\documentclass[11pt]{article}

\usepackage[margin=1in,letterpaper]{geometry}
\usepackage{fancyhdr}
\usepackage{amsmath,amssymb,amsthm}
\usepackage{graphicx}
\usepackage{booktabs}
\usepackage{multirow}
\usepackage{hyperref}
\usepackage{xcolor}
\usepackage{listings}
\usepackage{algorithm}
\usepackage{algpseudocode}
\usepackage{tikz}
\usetikzlibrary{shapes,arrows.meta,positioning,fit,backgrounds,calc}
\usepackage[expansion=false]{microtype}
\usepackage{caption}
\usepackage{subcaption}
\usepackage{placeins} % \FloatBarrier — keeps floats within their subsection
\usepackage{float}   % [H] placement specifier
\usepackage[numbers,sort&compress]{natbib}
\usepackage{url}
\usepackage{doi}
\usepackage{parskip} % space between paragraphs, no indent (arXiv style)
\newcommand{\chimax}{\chi_{\mathrm{max}}}
\newcommand{\Ct}{C(t)}
\newcommand{\ham}{\mathcal{H}}

\hypersetup{
 colorlinks=true,
 linkcolor=blue!60!black,
 citecolor=green!50!black,
 urlcolor=blue!60!black
}

\title{%
 \Large\bfseries GPU-First Heisenberg-Picture Tensor Network Dynamics\\
 for the 2D Transverse-Field Ising Model\\[0.5em]
}

\author{
 Paolo D'Alberto\thanks{Advanced Micro Devices, Inc. (AMD).
 \texttt{paolo.dalberto@amd.com}.
 This work was developed in active collaboration with Claude
 (Anthropic)}\\
 \small Advanced Micro Devices, Inc.\\
 \small\texttt{paolo.dalberto@amd.com}
}

\date{\today}

\begin{document}
\maketitle
\thispagestyle{fancy}

% ── Abstract ──────────────────────────────────────────────────────────────────
\begin{abstract}
We present \textsc{CppSim}, a C++/GPU 2D Ising simulator for
Heisenberg-picture tensor network time evolution on GPUs. The key
computational contributions are: first, a zero-malloc GPU workspace
that pre-allocates all buffers at startup; second, a custom GPU tensor
permutation kernel replacing host-side index shuffling with a pure
device-to-device operation, yielding a 7.6$\times$ trotter speedup;
third, a hybrid QR strategy selecting Cholesky-QR for tall-skinny
matrices and Householder-QR otherwise; fourth, adaptive Belief
Propagation with log-space Bethe partition function evaluation and
explicit sign tracking. The key validation contributions are: first,
correctness and demonstration of $\chi$-convergence of the spin
autocorrelation function $C(t)$ for $\chi \in \{20,\ldots,110\}$ on a
$4\times4$ grid, identifying $\chi^* \approx 60$ as the practical
convergence point. Second, we characterize BP fixed-point
multiplicity and introduce a four-step trap remediation procedure
using damped BP updates. Third, we identify operator-placement $C_4$
symmetry as the root cause of BP fixed-point multiplicity ---
symmetric grids are entirely trap-free. Fourth, grid scaling from
$3\times3$ to $10\times10$ demonstrates operator lightcone saturation
consistent with the Lieb--Robinson bound. A study across two 32\,GB GPU configurations reveals complementary BP
trap patterns and defines portable operating points at $\chi \in \{40, 60\}$.
An analytic memory model, validated to within 2\% against hardware
measurements, guided deployment on a 192+\,GB configuration, where we
demonstrate the first tensor-network simulation of the 2D Ising model on
a $10\times10$ lattice at $\chimax=100$ ($\approx109$\,GB) and an
$11\times11$ lattice at $\chimax=100$ ($\approx130$\,GB). The
$11\times11$ C4-symmetric run is entirely trap-free and reveals a clean
three-phase operator spreading: ballistic Lieb--Robinson cone, boundary
reflection ring, and onset of full scrambling.
\end{abstract}

%\tableofcontents
%\newpage

% ══════════════════════════════════════════════════════════════════════════════
\section{Introduction}
\label{sec:intro}
% ══════════════════════════════════════════════════════════════════════════════

Simulating real-time quantum dynamics of strongly correlated many-body systems is
a central challenge in computational physics. Tensor network (TN) methods provide
a systematically improvable approximation by representing quantum operators as
networks of low-rank tensors~\citep{orus2014practical}, with accuracy controlled
by the bond dimension~$\chi$. While one-dimensional systems are well-served by
matrix product state (MPS) methods~\citep{vidal2003efficient}, two-dimensional
(2D) tensor networks are increasingly important for studying frustrated magnets,
quantum circuits, and non-equilibrium dynamics at accessible system sizes.

The broader motivation for this work is the multi-product formula (MPF)
framework for hybrid quantum-classical computation. In MPF, a quantum computer
executes Trotter circuits to approximate Hamiltonian time evolution, and a
classical linear regression over multiple circuit runs at different step sizes
cancels the leading Trotter error terms. The spin autocorrelation function
$C(t) = \mathrm{Tr}[Z_c(t)\,Z_c]/2^N$ is the quantitative figure of merit:
it is what the ideal, error-free circuit would compute, and what the MPF
post-processing corrects toward. \textsc{CppSim} serves as a simulation sandbox: by computing the
ideal $C(t)$ alongside the Trotter-approximate trajectory, it enables
direct measurement of the circuit error and systematic exploration of
MPF compensation strategies before deployment on quantum hardware.
Throughout this work, all simulations are \emph{noise-free} ---
the only approximation is bond-dimension truncation; extension to
noisy gates and alternative lattice geometries (honeycomb, heavy-hex)
is left to future work.

The \emph{Heisenberg picture}~\citep{rudolph2025simulating} is the natural
language for this setting: the observable $Z_c$ is evolved forward in time
under the Hamiltonian while the infinite-temperature state remains fixed,
which maps directly to how a quantum circuit acts on an operator.
The correlation function is then extracted as an inner product with the
initial operator — no state preparation or measurement of a full density
matrix is required.
Computing $C(t)$ exactly would require contracting the full 2D tensor
network, an exponentially costly operation.
Belief propagation (BP)~\citep{tindall2023gauging} makes this tractable:
each lattice site iteratively sends its neighbors a message summarizing
everything it sees from its side of the network; after convergence, each
site holds an approximate picture of the global environment using only
local information.
The approximation rests on the Bethe assumption: on a tree, a site's
dependence on the rest of the network passes entirely through its direct
neighbors — there are no alternative paths — making BP exact.

Existing implementations of these methods are primarily in Julia~\citep{rudolph2025simulating},
benefiting from high-level tensor abstractions and just-in-time compilation. However,
Julia's garbage collector imposes non-trivial overhead for GPU workloads:
allocation patterns during simulation trigger frequent collection cycles, and GPU
memory pool management introduces latency spikes. For production-scale computations
at large bond dimensions and grid sizes, these overheads become significant.

We present \textsc{CppSim}, a C++/GPU implementation that addresses these limitations
through explicit GPU memory management, GPU-native linear algebra, and a zero-malloc
simulation loop, eliminating the memory pool and its management overhead entirely.
The implementation uses standard GPU programming interfaces and the BLAS/LAPACK
libraries~\citep{rocblas,rocsolver}, and introduces custom device-to-device tensor
permutation kernels to eliminate CPU--GPU traffic entirely, adaptive QR strategies
to exploit matrix shapes, and a sign-correct log-space Bethe partition function
with explicit edge correction.

Because all buffer sizes are determined analytically, \textsc{CppSim}
pre-allocates the minimum allocation that sustains maximum parallelism
before the first layer runs: every intermediate tensor occupies a named,
fixed slot and no dynamic allocation occurs during simulation.
This is not conservative over-provisioning --- the four $O(\chimax^4)$
buffers (two contracted site tensors and two QR factors) must coexist at
every gate and set an irreducible lower bound; no smaller pre-allocation
could support full parallel execution at $\chimax$.
The resulting closed-form memory model (\texttt{estimate\_memory.py})
predicts VRAM requirements for any grid and bond dimension to within 2\%
of hardware measurements, enabling deployment decisions before a single
simulation step is taken.

The remainder of this paper is organized as follows. Section~\ref{sec:physics}
describes the mathematical formulation. Section~\ref{sec:design} presents the
software design. Section~\ref{sec:algorithms} details the core algorithms.
Section~\ref{sec:implementation} describes the implementation.
Section~\ref{sec:experiments} presents numerical experiments and physics results.
Section~\ref{sec:performance} reports performance. Section~\ref{sec:portability}
analyses the memory-bandwidth regime and numerical precision. Section~\ref{sec:conclusion} concludes.

% ══════════════════════════════════════════════════════════════════════════════
\section{Mathematical Formulation}
\label{sec:physics}
% ══════════════════════════════════════════════════════════════════════════════

This section establishes the physical model and the mathematical framework
underlying the simulation. We work entirely in the Heisenberg picture, where
the observable rather than the quantum state is evolved in time, and we
represent it as a tensor network on the lattice graph.

\subsection{Model}

We consider the 2D transverse-field Ising model~\citep{sachdev2011quantum} on a
$k \times l$ rectangular lattice with open boundary conditions and Hamiltonian:
\begin{equation}
 \ham = -J \sum_{\langle ij \rangle} X_i X_j \;-\; h \sum_i Z_i,
 \label{eq:hamiltonian}
\end{equation}
where $X_i, Z_i$ are Pauli operators on site $i$, $J$ is the Ising coupling,
$h$ is the transverse field strength, and the sum is over nearest-neighbor pairs.

\subsection{Heisenberg Picture and Spin Autocorrelation}

The spin autocorrelation function measures how much the observable $Z_c$
at the lattice center remembers its initial value after time $t$:
\begin{equation}
 C(t) = \mathrm{Tr}\bigl[\rho_0\, Z_c(t)\, Z_c\bigr],
 \label{eq:autocorr_general}
\end{equation}
where $\rho_0$ is the initial state.
In the Heisenberg picture, the state $\rho_0$ remains fixed while the
observable evolves under the Hamiltonian:
\begin{equation}
 Z_c(t) = e^{i\ham t}\, Z_c\, e^{-i\ham t}.
 \label{eq:heisenberg}
\end{equation}
Choosing $\rho_0$ constant (proportional to the identity), the trace
simplifies to:
\begin{equation}
 C(t) = \frac{1}{2^N} \mathrm{Tr}\bigl[ Z_c(t)\, Z_c \bigr],
 \label{eq:autocorr}
\end{equation}
where $2^N$ is the Hilbert space dimension ($N$ sites),
which is the quantity computed throughout this work and targeted by the
MPF post-processing described in Section~\ref{sec:intro}.

\subsection{Pauli Basis Tensor Network}

Each qubit site has a 2-dimensional Hilbert space, but the space of $2\times2$
Hermitian operators has dimension $D=4$, spanned by the Pauli matrices
$\{I, X, Y, Z\}$ — orthogonal under the Hilbert-Schmidt inner product
$\mathrm{Tr}(A^\dagger B)$.
The operator $Z_c(t)$ is represented as a tensor network on the lattice graph:
each site $v$ carries a tensor $\psi_v$ whose indices encode both the bond
entanglement with neighboring sites and the local Pauli component of the operator.

Concretely, $\deg(v)$ denotes the number of nearest-neighbor bonds of site $v$:
2 at corners, 3 at edge sites, and 4 at interior sites on a rectangular grid.
The tensor $\psi_v$ is pre-allocated with one bond index of dimension $\chimax$
per neighbor and one Pauli index of dimension $D=4$, giving a maximum shape of
$\chimax^{\deg(v)} \times D$. The live bond dimension $\chi_{\mathrm{live}}$
grows from 1 to $\chimax$ as entanglement builds across the first Trotter layers;
all computations use the current live dimension, not the pre-allocated maximum.
The entry $\psi_v[i_1, \ldots, i_{\deg(v)},\, p]$ is the coefficient of the
$p$-th Pauli matrix ($p{=}0$: $I$,\ $p{=}1$: $X$,\ $p{=}2$: $Y$,\ $p{=}3$: $Z$)
at site $v$, for bond state $(i_1,\ldots,i_{\deg(v)})$.
At $t=0$: the center site has $\psi_c[\cdot,3]=1$ (pure $Z$); all others have
$\psi_v[\cdot,0]=1$ (pure $I$).

The bond entanglement between neighboring sites $a$ and $b$ is captured by one
$\chi \times \chi$ message matrix $\mu_{a \to b}$ per directed bond.
During Belief Propagation, each $\mu_{a \to b}$ is a general
positive-semidefinite matrix updated iteratively until a self-consistent
fixed point is reached.
At that fixed point the matrices become diagonal, and their diagonal entries
are the \emph{Schmidt singular values} of the bipartition of the network
across bond $(a,b)$: each entry $\lambda_k \geq 0$ is the weight of one
entanglement channel across the cut, in the same way that the singular values
of a matrix decompose it into rank-one pieces ordered by importance.
A product state (no entanglement) has only $\lambda_1 > 0$; a maximally
entangled bond has all $\chi$ values equal.
This fixed point is the \emph{Vidal canonical gauge}~\citep{vidal2004efficient},
in which the site tensors are orthogonalized so that all entanglement
information is encoded in the diagonal messages, and truncating each bond to
rank $\chimax$ discards only the least significant channels — the locally
optimal low-rank approximation at fixed bond dimension.
On a tree this gauge is exact; on the 2D lattice BP finds an approximate
Vidal gauge, which is why BP runs before every correlation function evaluation.

\subsection{Trotter Decomposition}

Each time step $\delta t$ advances the operator by unitary conjugation:
\begin{equation}
 Z_c(t+\delta t) = U^\dagger(\delta t)\; Z_c(t)\; U(\delta t),
 \qquad U(\delta t) = e^{-i\ham\delta t},
 \label{eq:heisenberg_step}
\end{equation}
where $U^\dagger = e^{+i\ham\delta t}$ acts on the left and $U$ on the right.
Because the Hamiltonian~\eqref{eq:hamiltonian} contains only real Pauli couplings,
this conjugation maps Pauli matrices to real linear combinations of Pauli matrices,
so the gate acting on $\psi_v$ in the Pauli basis is a real orthogonal matrix —
no complex arithmetic is required.

The unitary $U(\delta t)$ is itself approximated to first order as a
composition of simpler gates (the Trotter
decomposition~\citep{suzuki1976generalized}):
\begin{equation}
 U(\delta t) \approx R_z(h\,\delta t) \cdot R_{xx}(2J\,\delta t) \cdot R_z(h\,\delta t),
 \label{eq:trotter}
\end{equation}
where $R_z$ is a single-site rotation (diagonal in the Pauli basis) and $R_{xx}$
is a two-site entangling gate applied to all nearest-neighbor pairs via a
4-color decomposition: bonds are partitioned into four color classes such that no
two bonds in the same class share a site, enabling sequential independent updates.

\subsection{Gate Update}
\label{sec:gate_update}

For each two-site gate on bond $(v_1, v_2)$, let $i \in \{1,2\}$ index the two
sites. The message matrices $\mu_{a\to b}$ (defined in \S\ref{sec:physics}) are
diagonal in the Vidal gauge, holding singular values; for the bond under
update, $\sqrt{\mu_{v_1\to v_2}}$ and $\sqrt{\mu_{v_2\to v_1}}$ denote the
diagonal matrices of their square roots, and $\sqrt{\mu_{e\to v_i}}$ denotes
the same for every other incident bond $e$.
The update proceeds as follows (see Algorithm~\ref{alg:gate}):
\begin{enumerate}
 \item \textbf{Environment absorption.} Contract $\psi_{v_i}$ with
 $\sqrt{\mu_{e \to v_i}}$ on every bond $e$ incident to $v_i$ except the
 active bond $(v_1,v_2)$. This yields the \emph{contraction vector}
 $\mathbf{cv}_i$ — so named because it is the site tensor contracted
 with its environment, ready for QR — which folds the environmental
 context into each site before the update.
 \item \textbf{QR decomposition.} Factor each gauged tensor:
 $\mathbf{cv}_i = \mathbf{Q}_i \mathbf{R}_i$,
 where $\mathbf{Q}_i$ ($\chimax^3 \times D\chimax$) is isometric and
 $\mathbf{R}_i$ ($D\chimax \times D\chimax$) is compact.
 QR reduces the subsequent SVD from the intractable full tensor
 ($\chimax^3 \times D\chimax$) to a small bond matrix ($D\chimax \times D\chimax$).
 \item \textbf{Bond matrix.} Form $\mathbf{RR} = \mathbf{R}_1^\top \mathbf{R}_2$
 ($D\chimax \times D\chimax$), coupling the two sites at the bond interface.
 \item \textbf{Gate contraction.} The gate $G \in \mathbb{R}^{D^2 \times D^2}$
 acts on the $D \times D$ physical (Pauli) block of $\mathbf{RR}$,
 leaving the bond indices unchanged:
 \[
 \tilde{R}[d_1',d_2',k,l]
 = \sum_{d_1,d_2} G[d_1'd_2',\,d_1d_2]\;\mathbf{RR}[d_1,d_2,k,l].
 \]
 \item \textbf{SVD and truncation.} Factor
 $\tilde{\mathbf{R}} = \mathbf{U}\mathbf{S}\mathbf{V}^\top$,
 truncated to rank $\chimax$ (equation~\eqref{eq:truncation}).
 \item \textbf{Reconstruction.} Set $\mathbf{cv}_i^{\text{new}} = \mathbf{Q}_i \mathbf{M}_i$,
 where $\mathbf{M}_1 = \mathbf{U}\sqrt{\mathbf{S}}$ and
 $\mathbf{M}_2 = \sqrt{\mathbf{S}}\mathbf{V}^\top$.
 \item \textbf{Deabsorption.} Multiply by $(\sqrt{\mu_{e\to v_i}})^{-1}$ on all
 environment bonds to restore the Vidal gauge.
 \item \textbf{Vidal normalization.} Store
 $\lambda_k \leftarrow s_k/\|\mathbf{s}\|_2$ ($\sum_k \lambda_k^2 = 1$),
 preventing float32 overflow at large $\chi$.
\end{enumerate}

\subsection{Correlation Function via Belief Propagation}

$C(t)$~\eqref{eq:autocorr} is the normalized Hilbert-Schmidt inner product of
the evolved operator $Z_c(t)$ with the initial operator $Z_c$:
$C(t) = \langle Z_c(t),\, Z_c \rangle_{HS}$, where
$\langle A, B\rangle_{HS} = \mathrm{Tr}[A^\dagger B]/2^N$.
Evaluating it requires contracting the full tensor network against the observable —
a three-step process.

\textbf{Step 1: trace to tensor network contraction.}
By Pauli orthogonality ($\mathrm{Tr}[P_a P_b] = 2\delta_{ab}$), the trace
reduces to a single tensor network contraction: the coefficient of the
configuration ``$Z$ at center $c$, $I$ at all other sites'' in the full network.
Concretely, each site tensor $\psi_v$ has shape $[\chi,\chi,\chi,\chi,4]$ ---
four bond legs plus one physical leg of size 4 (one entry per Pauli $\{I,X,Y,Z\}$).
Pauli orthogonality pins each physical leg to a single value: $Z$ at site $c$,
$I$ at every other site, reducing each tensor to $[\chi,\chi,\chi,\chi]$.
The remaining bond legs are shared between neighboring sites; contracting them ---
summing over every shared bond index across the whole lattice --- yields a single
scalar, which is $C(t)$.

\textbf{Step 2: Bethe factorization.}
Contracting a 2D tensor network exactly is exponentially costly.
After BP has converged the messages to the approximate Vidal gauge, the
Bethe approximation (treating the lattice as a tree) factorizes the contraction
as a product over sites divided by a product over edges:
\begin{equation}
 C(t) = \frac{\displaystyle\prod_v vs_v}{\displaystyle\prod_e Z_e},
 \label{eq:bethe_raw}
\end{equation}
where $vs_v$ is the scalar contraction of $\psi_v$ with its BP environment and
the observable ($Z$ at center $c$, $I$ elsewhere), and
\begin{equation}
 Z_e = \sum_i \mu_{a \to b}[i]\, \mu_{b \to a}[i]
 \label{eq:ze}
\end{equation}
is the Bethe edge correction: each bond appears in two site contractions, and
$Z_e$ removes the double-counting.

\textbf{Step 3: log-space evaluation and sign tracking.}
The $Z$ operator has eigenvalues $\pm 1$, so the observable-specific BP messages
can be negative, making $vs_v$ and $Z_e$ real-valued but not necessarily positive.
Taking absolute values and separating signs before applying logarithms gives the
numerically stable form:
\begin{equation}
 C(t) = \exp\!\Bigl(\sum_v \log|vs_v| \;-\; \sum_e \log|Z_e|\Bigr)
 \times \underbrace{\prod_v \mathrm{sign}(vs_v)}_{\mathrm{sign\_num}}
 \times \underbrace{\prod_e \mathrm{sign}(Z_e)}_{\mathrm{sign\_den}},
 \label{eq:bethe}
\end{equation}
Omitting the sign factors would flip the sign of $C(t)$ whenever the product
of signs across all sites and edges is $-1$ — a catastrophic error at late times
when the sign structure of the BP messages becomes non-trivial.

% ══════════════════════════════════════════════════════════════════════════════
\section{Software Design}
\label{sec:design}
% ══════════════════════════════════════════════════════════════════════════════

The overarching design principle is that GPU memory allocation is a one-time
startup cost. Every buffer needed by the simulation --- site tensors, BP messages,
workspace intermediates --- is allocated at initialization at its worst-case size
(i.e.,\ at $\chimax$) and reused for every gate, layer, and Trotter step.
This is the minimum allocation that sustains maximum parallelism: the four
$O(\chimax^4)$ tensors that must coexist during every gate application set
an irreducible lower bound, and no smaller pre-allocation could support full
parallel execution at $\chimax$.
This section describes the two main structures that realize this principle
and the sizing rules that must be followed exactly.

\subsection{Zero-Malloc Simulation Loop}

The central design principle is that \textbf{no GPU memory is allocated or freed
during simulation}. All buffers are allocated once at startup in two structures:

\paragraph{Workspace} (grid-independent, reused for every gate):
\begin{center}
\begin{tabular}{@{}lll@{}}
\toprule
Buffers & Size & Purpose \\
\midrule
\texttt{cv1, cv2, Q1, Q2} & $\chimax^3 \times D\chimax$ & gauged tensors, QR factors \\
\texttt{R1, R2, Rinv1, Rinv2,} & \multirow{2}{*}{$(D\chimax)^2$ or $D\chimax$} & \multirow{2}{*}{QR/SVD matrices} \\
\texttt{RR, U, Vt, gram, S} & & \\
\texttt{sqrt\_env[4], inv\_sqrt\_env[4]} & $\chimax^2$ & environment absorption \\
\texttt{svd\_work} & $D^2\chimax$ & GESVD bidiagonal scratch \\
\bottomrule
\end{tabular}
\end{center}

\paragraph{IsingState} (grid-dependent, persistent across all layers):
\begin{center}
\begin{tabular}{@{}lll@{}}
\toprule
Buffers & Size & Purpose \\
\midrule
\texttt{psi[}$v$\texttt{]} & $D \cdot \chimax^{\deg(v)}$ & site tensors, pre-allocated at $\chimax$ \\
\texttt{messages[bond][dir]} & $\chimax^2$ & Vidal gauge, one per directed bond \\
\bottomrule
\end{tabular}
\end{center}

Peak VRAM is reached at startup and remains constant throughout the run,
eliminating GC pressure, fragmentation, and allocator contention.

\subsection{Buffer Sizing and Safety Margins}

The SVD bidiagonal scratch buffer (\texttt{svd\_work}) requires careful sizing.
The SVD is called on a matrix of dimensions $k_{1D} \times k_{2D}$ where
$k_{iD} = k_i \times D$ and $k_i = \min(m_i, n_i)$. At bond dimension saturation,
$m_i = \chimax^3 \gg n_i = \chimax D$, so $k_i = \chimax D$ and
$k_{iD} = \chimax D^2$. The LAPACK \texttt{SGESVD} bidiagonal scratch (\texttt{E})
must therefore be sized to $\min(k_{1D}, k_{2D}) = \chimax D^2$ elements — a
factor of $D=4$ larger than the naive estimate $\chimax D$.

Both site tensors $\psi_v$ and message matrices $\mu_{a\to b}$ are allocated at their maximum possible size from
startup, even when the actual bond dimension is much smaller at early layers.
This trades $O(\chimax^4)$ memory at early layers for zero reallocation.
The actual computation uses live \texttt{bond.chi} values, which grow from 1 to
$\chimax$ over the first few Trotter layers.
Because leading dimensions are fixed at startup, all buffers can in principle be
padded to alignment boundaries (multiples of the HBM cache-line width) at
allocation time; once bond dimensions reach $\chimax$ at saturation, this
guarantees coalesced memory access at every \textsc{BLAS} call ---
an optimization that is unavailable to allocators that resize buffers
dynamically per layer.

% ══════════════════════════════════════════════════════════════════════════════
\section{Algorithms}
\label{sec:algorithms}
% ══════════════════════════════════════════════════════════════════════════════

\textsc{CppSim} was written entirely from scratch in C++/GPU, without reusing
any code from the Julia reference. The physics, tensor operations, and BP
were all re-implemented, and each required algorithmic choices that are specific
to the GPU execution model and not present in the Julia implementation.
Four components in particular represent original contributions: a GPU-native
tensor permutation kernel that eliminates PCIe traffic entirely, an adaptive
QR strategy exploiting the tall-thin structure of the gauged contraction vectors,
a truncated SVD on the compact bond matrix enabled by that QR, and a
log-space Bethe partition function with explicit sign tracking that resolves
systematic late-time errors. We describe each in turn.

\subsection{GPU Tensor Permutation}
\label{sec:permute}

Site tensors $\psi_v$ are rank-5 objects (for interior sites) stored in canonical
bond order $[\text{U}, \text{D}, \text{L}, \text{R}, \text{Pauli}]$,
where the Pauli index (dimension $D=4$) selects one of the four matrices
$\{I, X, Y, Z\}$ — each a $2\times2$ matrix with entries in $\{0, \pm1\}$ (or
$\pm i$ for $Y$).

QR decomposition operates on matrices (rank-2 tensors), so $\psi_v$ must first
be reshaped into one. The natural split is: group the \emph{environment} bond
indices (all bonds except the one being updated, dimension $\chimax^3$ for an
interior site) into the rows, and group the \emph{cut bond} together with the
Pauli index (dimension $\chimax \times D = D\chimax$) into the columns.
This produces a tall-thin matrix of shape $\chimax^3 \times D\chimax$, from
which QR extracts an isometric $\mathbf{Q}$ (the environment) and a compact
$\mathbf{R}$ (the bond interface).
To enable this grouping, the cut-bond index must be moved adjacent to the Pauli
index before reshaping — that is the sole purpose of the tensor permutation.

Permuting a rank-5 tensor of $\chimax^4 D$ elements ---
the complete site tensor $\Psi_v$, carrying the full bond-index state of a
single lattice site --- on the CPU requires:
a device-to-host (D2H) transfer ($\sim$100\,MB per interior site at $\chi=50$),
an index-remapping loop, and a host-to-device (H2D) transfer back ---
totaling $\sim$9.6\,GB of PCIe traffic per Trotter layer at $\chi=50$.

Our custom GPU kernel eliminates all PCIe traffic. Each thread handles one output
element by decomposing the input flat index (column-major) into a multi-index,
applying the permutation, and computing the output flat index:

\begin{lstlisting}[language=C++, basicstyle=\ttfamily\scriptsize,
 caption={GPU tensor permutation kernel}, float=ht]
__global__ void tensor_permute_kernel(
 const float* __restrict__ src, // input tensor (column-major flat)
 float* __restrict__ dst,        // output tensor (permuted flat)
 PermuteMeta meta, int total)    // shape, strides, rank; passed by value
{
 int idx = blockIdx.x * blockDim.x + threadIdx.x; // one thread per element
 if (idx >= total) return;

 int in_idx = idx, out_idx = 0;
 for (int d = 0; d < meta.ndim; d++) {
   int coord = in_idx % meta.shape_in[d]; // coordinate along dim d
   in_idx   /= meta.shape_in[d];          // peel off dim d
   out_idx  += coord * meta.stride_out[d];// accumulate permuted offset
 }
 dst[out_idx] = src[idx]; // scatter to permuted position
}
\end{lstlisting}

The \texttt{PermuteMeta} struct (shape, output strides, permutation, rank) is
passed by value to avoid a device pointer indirection. This kernel reduced trotter
time from $\sim$49{,}000\,ms/layer to $\sim$6{,}400\,ms/layer at $\chi=50$
saturation — a \textbf{7.6$\times$ speedup} from this single change.

The permutation itself is conceptually straightforward: on the CPU it reduces
to a simple index-remapping loop, easily expressed in any language.
However, a device-side implementation is not readily available in high-level
GPU frameworks such as Julia/GPU: although tensor index permutation is a
frequent and performance-critical operation in tensor network codes, an
optimized on-device kernel is rarely provided out of the box, leaving
frameworks to fall back on D2H--remap--H2D round-trips.
The kernel presented here is therefore a self-contained, reusable contribution:
dropped into a Julia/GPU tensor network implementation it would deliver the
same PCIe-traffic elimination and comparable speedup, independently of the rest
of \textsc{CppSim}.

\subsection{Adaptive QR Decomposition}
\label{sec:qr}

At each gate application, a QR decomposition of the gauged tensor
$\mathbf{cv} \in \mathbb{R}^{m \times n}$ is required, where
$m = \chimax^3$ and $n = D\chimax$ at saturation. Two paths are used:

\paragraph{Cholesky-QR~\citep{golub2013matrix}} (when $n > 150$ and $m/n > 150$;
thresholds left as tunable parameters since the optimal crossover
depends on both matrix dimensions and memory bandwidth):
\begin{equation}
 \mathbf{C} = \mathbf{cv}^\top \mathbf{cv} + \epsilon \mathbf{I}, \quad
 \mathbf{R} = \mathrm{chol}(\mathbf{C}), \quad
 \mathbf{Q} = \mathbf{cv}\, \mathbf{R}^{-1},
 \label{eq:cholqr}
\end{equation}
with $\epsilon = 10^{-8}$ ensuring positive definiteness. This path uses
BLAS \texttt{SGEMM} for $\mathbf{C}$, LAPACK \texttt{SPOTRF} for the
Cholesky factor, and BLAS \texttt{STRSM} to solve
$\mathbf{Q}\,\mathbf{R} = \mathbf{cv}$ entirely on-device.
The only host transfer is a single integer (\texttt{potrf\_info}, 4 bytes)
to confirm that $\mathbf{C}$ was positive definite; all matrix operations
remain on the GPU.

\paragraph{Householder-QR} (primary path at early layers; fallback at saturation):
\[
 (\mathbf{Q}, \mathbf{R}) = \mathrm{qr}(\mathbf{cv}),
\]
via LAPACK \texttt{SGEQRF} + \texttt{SORGQR}. This path is backward stable
and numerically safe regardless of the condition number of $\mathbf{cv}$,
since it works directly on the matrix without forming the gram matrix.
Both paths are communication-free for their matrix operations.
In the Householder path, \texttt{SGEQRF} stores the result
in-place: the Householder reflectors occupy the lower triangle and $\mathbf{R}$
sits in the upper triangle of the same buffer. A custom on-device kernel
(\texttt{extract\_upper\_triangle}) copies $\mathbf{R}$ into its own buffer
without any host transfer, after which \texttt{SORGQR} generates $\mathbf{Q}$
on-device from the reflectors.
At early layers ($n \leq 150$), $\chi$ is still small so the matrices are tiny.
When invoked as a fallback at saturation, the same buffer is
$\chimax^3 \times D\chimax \approx 1.6$\,GB at $\chi=100$ --- a case that
never arises in practice.\footnote{In the Julia
reference implementation, we instrumented the equivalent adaptive QR path and
measured zero fallback invocations across all runs at all bond dimensions.
Since the fallback would have been triggered by either a Cholesky failure
or a numerical anomaly in the result, its absence confirms both: the
Cholesky-QR path is stable in practice for this problem class, with no
sign of numerical degradation --- neither catastrophic nor mild.}

\paragraph{Numerical precision and the Gram matrix accumulation error.}
The Cholesky-QR path involves two sources of numerical difficulty that compound
at high bond dimension.
First, forming $\mathbf{C} = \mathbf{cv}^\top \mathbf{cv}$ via GEMM accumulates
$m = \chimax^3$ floating-point products per entry; at $\chi=100$ this gives
$m \approx 10^6$ terms, so the float32 rounding floor
($\varepsilon \approx 10^{-7}$) contributes an absolute error of order
$m\varepsilon \approx 10^{-1}$ per entry of $\mathbf{C}$---enough to corrupt
near-zero eigenvalues and destroy positive-definiteness.
Second, $\kappa(\mathbf{C}) = \kappa(\mathbf{cv})^2$, so any mild ill-conditioning
in the input is squared in the Gram matrix.
Both effects are alleviated by computing the gate Trotter expansion in float64:
the accumulation floor drops to $m\varepsilon_{64} \approx 10^{-10}$ and the
input matrix is better conditioned, making $\mathbf{C}$ reliably positive-definite
without the $\epsilon\mathbf{I}$ regulariser.
Mixed-precision GEMM (\texttt{GEMM-ex} with float64 accumulators) or a
dedicated \texttt{DGEMM}+\texttt{DPOTRF} path would deliver float64 Gram
matrix formation with no change to the float32 tensor storage, eliminating
the Householder fallback entirely and allowing Adaptive QR to be used
unconditionally.
Even in its current float32 form, the experimental results demonstrate the
right algorithmic direction: on bandwidth-bound configurations Adaptive QR
matches Householder, confirming that Cholesky's lower compute constant is
masked by memory traffic; on the high-bandwidth configuration, where data
arrives faster than compute can consume it, Adaptive QR delivers a 22--35\%
speedup --- exactly the regime where Cholesky's arithmetic advantage becomes
visible (Section~\ref{sec:portability} and Appendix~\ref{app:mi355x}).

\paragraph{SVD and Truncation.}
The bond matrix $\mathbf{RR} \in \mathbb{R}^{k_{1D} \times k_{2D}}$ is factored
via LAPACK \texttt{SGESVD}. The truncation rank is:
\begin{equation}
 \chi_{\text{new}} = \min\!\bigl(\#\{i : s_i \geq \epsilon_{\text{cut}}\, s_0\},\;
 \chimax\bigr),
 \label{eq:truncation}
\end{equation}
with default cutoff $\epsilon_{\text{cut}} = 10^{-11}$.
Singular values are Vidal-normalized before storage:
\begin{equation}
 \lambda_i = s_i \,/\, \bigl(\textstyle\sum_j s_j^2\bigr)^{1/2},
 \qquad \sum_i \lambda_i^2 = 1.
 \label{eq:vidal_norm}
\end{equation}
By construction, $\sum_k \lambda_k^2 = 1$ implies $\lambda_k \in [0,1]$ for
any $\chi$ — float32 overflow of the message entries is impossible regardless
of bond dimension. Without this normalization, overflow was observed in practice
at $\chi \geq 50$ saturation.

\paragraph{BP Message Update.}
Each BP sweep updates all messages via:
\begin{equation}
 \mu_{v \to e}^{\text{new}}[i] \propto
 \psi_v[i, \ldots] \cdot \prod_{e' \neq e} \mu_{e' \to v}[i],
 \label{eq:bp}
\end{equation}
implemented via BLAS strided-batched \texttt{SGEMV}.
Normalization is entirely on-device: \texttt{SDOT} in
\texttt{POINTER\_MODE\_DEVICE} writes the sum directly to a device scalar
\texttt{d\_s}, which the \texttt{normalize\_and\_delta} kernel reads to rescale
the message in the same pass --- no host transfer required.
The convergence delta is accumulated on-device via \texttt{atomicMax} across
all bond updates in the sweep, producing a single device float \texttt{d\_delta}.
One D2H transfer of 4 bytes occurs per sweep (not per message) to read
\texttt{d\_delta} and make the stop/continue decision --- unavoidable, since
loop control requires a host-visible value.

Adaptive convergence: the sweep loop exits when
\begin{equation}
 \max_i |\mu^{\text{new}}[i,i] - \mu^{\text{old}}[i,i]| < \tau = 10^{-5},
 \label{eq:bp_tol}
\end{equation}
returning the sweep count. Most layers converge in 3--20 sweeps; hard layers
(typically alternating layers at late times) require the full 100-sweep budget.

\begin{algorithm}
\caption{Gate update for bond $(v_1, v_2)$.}
\label{alg:gate}
\begin{algorithmic}[1]
\Require Site tensors $\psi_{v_1}, \psi_{v_2}$, gate $G$, messages $\mu$, $\chimax$
\Ensure Updated $\psi_{v_1}, \psi_{v_2}$, updated message $\mu_{v_1 \leftrightarrow v_2}$
\State Compute $\sqrt{\mu}$, $(\sqrt{\mu})^{-1}$ via \textsc{Syevd} for each env bond
\State $\mathbf{cv}_i \leftarrow \textsc{AbsorbEnv}(\psi_{v_i},\, \sqrt{\mu})$
 \quad $i = 1, 2$
\State $\mathbf{cv}_i \leftarrow \textsc{PermuteCutToLast}(\mathbf{cv}_i)$
\State $[\mathbf{Q}_i, \mathbf{R}_i] \leftarrow \textsc{AdaptiveQR}(\mathbf{cv}_i)$
\State $\mathbf{RR} \leftarrow \mathbf{R}_1^\top \mathbf{R}_2$
 \quad (via \textsc{Sgemm})
\State $\tilde{\mathbf{R}} \leftarrow G(\mathbf{RR})$
 \quad (host loop over $D^4$ entries)
\State $[\mathbf{U}, \mathbf{s}, \mathbf{V}^\top] \leftarrow \textsc{Sgesvd}(\tilde{\mathbf{R}})$
\State $\chi_{\text{new}} \leftarrow \min(\#\{s_i \geq \epsilon_{\text{cut}} s_0\},\, \chimax)$
\State $\mathbf{cv}_i^{\text{new}} \leftarrow \mathbf{Q}_i \mathbf{M}_i$
 \quad (bridge matrix $\mathbf{M}_i$ from $\mathbf{U}/\mathbf{V}^\top$, $\sqrt{\mathbf{s}}$)
\State $\mathbf{cv}_i^{\text{new}} \leftarrow \textsc{InversePermuteCut}(\mathbf{cv}_i^{\text{new}})$
\State $\mathbf{cv}_i^{\text{new}} \leftarrow \textsc{DeabsorbEnv}(\mathbf{cv}_i^{\text{new}},\,
 (\sqrt{\mu})^{-1})$
\State $\psi_{v_i} \leftarrow \mathbf{cv}_i^{\text{new}}$
\State $\mu_{v_1 \to v_2} \leftarrow \mu_{v_2 \to v_1}
 \leftarrow \mathrm{diag}(\lambda)$,
 \quad $\lambda_i = s_i/\|\mathbf{s}\|_2$
\end{algorithmic}
\end{algorithm}

% ══════════════════════════════════════════════════════════════════════════════
\section{Implementation}
\label{sec:implementation}
% ══════════════════════════════════════════════════════════════════════════════

% ── Pipeline figure ───────────────────────────────────────────────────────────
% \sw{calls} formats the software-call row at the bottom of each pipeline box.
\newcommand{\sw}[1]{\\[1pt]{\color{gray!55}\footnotesize\ttfamily\hspace{1em}#1}}
\begin{figure}[p]
\centering
\resizebox{\textwidth}{0.93\textheight}{%
\begin{tikzpicture}[
 font=\small,
 every node/.style={align=left},
 outerbox/.style={draw, rounded corners=5pt, thick,
 text width=13.8cm, minimum height=0.65cm,
 inner xsep=7pt, inner ysep=3pt,
 fill=#1!12, draw=#1!70!black},
 stepbox/.style={draw, rounded corners=3pt,
 text width=13.2cm, minimum height=0.56cm,
 inner xsep=7pt, inner ysep=3pt,
 fill=#1!10, draw=#1!60!black, font=\small},
 arr/.style={-{Stealth[length=4pt,width=3pt]}, thick, gray!70},
 groupfit/.style={draw, rounded corners=6pt, dashed, thick,
 draw=gray!50, inner sep=7pt},
 looplabel/.style={font=\footnotesize\itshape, text=gray!70, align=center},
]

% ── Outer loop header ─────────────────────────────────────────────────────────
\node[outerbox=gray] (init)
 {Initialize:\ $\psi_c \!\gets\! Z$,\ $\psi_{v\neq c}\!\gets\! I$,\
 $\mu \!\gets\! \mathbf{I}_1$\ (all bonds $\chi\!=\!1$)};

\node[looplabel, below=0.14cm of init] (forloop)
 {for $t = 1,\ldots,T$:};

% ── Phase 1: Trotter ──────────────────────────────────────────────────────────
\node[outerbox=blue, below=0.14cm of forloop] (rz1)
 {$R_z(h\,\delta t)$: apply unary phase gate to all sites%
 \sw{apply\_unary\_gate}};

\node[looplabel, below=0.35cm of rz1] (colorloop)
 {4-color $R_{xx}$ sweep\ —\ for each gate $(v_1,\,v_2)$:};

% inner gate steps
\node[stepbox=teal, below=0.14cm of colorloop] (psqrt)
 {\textbf{1.}\ \textsc{pseudo\_sqrt}\ per env bond of $v_1$ (then $v_2$):\
 symmetrize $\mu$ $\to$ syevd $\to$ form
 $\sqrt{\mu}$ and $\mu^{-1/2}$ on-device%
 \sw{symmetrize\_copy\quad SSYEVD\quad form\_sqrt\_matrices}};

\node[stepbox=teal, below=0.07cm of psqrt] (absorb)
 {\textbf{2.}\ Absorb env, per site $i\!\in\!\{1,2\}$:\
 $\mathbf{cv}_i \gets \sqrt{\mu_{e\to v_i}}\cdot\psi_{v_i}$
 on all non-cut neighbor bonds%
 \sw{SGEMM-batched}};

\node[stepbox=blue!70!black, below=0.07cm of absorb] (perm1)
 {\textbf{3.}\ Permute cut bond to last index, per site\ (no-op if already last)%
 \sw{gpu\_tensor\_permute}};

\node[stepbox=blue!70!black, below=0.07cm of perm1] (qr)
 {\textbf{4.}\ QR $\times 2$, per site:\ $\mathbf{cv}_i = \mathbf{Q}_i\mathbf{R}_i$;\quad
 $n_i = \chi_{\mathrm{cut}}D$,\
 $m_i = \prod_{e\neq\mathrm{cut}}\chi_e$\ \ (sizes differ: corner $m\!=\!\chi D$,
 edge $m\!=\!\chi^2 D$, interior $m\!=\!\chi^3 D$)%
 \sw{SPOTRF + STRSM\ \ (Cholesky)\ /\ SGEQRF + SORGQR\ \ (Householder)}};

\node[stepbox=blue!70!black, below=0.07cm of qr] (rr4)
 {\textbf{5.}\ Reshape $R_i\!\to\!A_i$: separate $\chi_{\mathrm{cut}}$ and $D$ indices ($\times 2$);\quad
 form $\mathbf{RR4} = \mathbf{A}_1 \mathbf{A}_2^\top\ \
 [k_1 D \times k_2 D]$%
 \sw{reshape\_R\_to\_A\quad (kernel, $\times 2$)\qquad SGEMM}};

\node[stepbox=orange!90!black, below=0.07cm of rr4] (gate)
 {\textbf{6.}\ Contract gate:\ $\mathbf{oR4}_{d_1'd_2'kl}
 = \sum_{d_1d_2} G_{d_1'd_2',d_1d_2}\,\mathbf{RR4}_{d_1d_2kl}$%
 \sw{apply\_gate\_to\_RR4\quad (kernel, shared-mem $G$\,[$D^4$ floats])}};

\node[stepbox=orange!90!black, below=0.07cm of gate] (svd)
 {\textbf{7.}\ SVD:\ $\mathbf{oR4} = \mathbf{U}\mathbf{S}\mathbf{V}^\top$;\quad
 truncate to $\chi_{\mathrm{new}} \le \chimax$;\quad
 Vidal normalize: $\lambda_k \gets s_k/\|\mathbf{s}\|_2$\quad
 [\,only CPU step: D2H of $k_{\mathrm{full}}$ singular values\,]%
 \sw{SGESVD}};

\node[stepbox=red!70!black, below=0.07cm of svd] (recon)
 {\textbf{8.}\ Build $\mathbf{M}_1$ from $\mathbf{U}\cdot\sqrt{S}$
 and $\mathbf{M}_2$ from $\mathbf{V}^\top\!\cdot\!\sqrt{S}$,
 per site ($\times 2$);\quad
 reconstruct $\mathbf{cv}_i = \mathbf{Q}_i\mathbf{M}_i$%
 \sw{build\_M\_from\_U\quad build\_M\_from\_Vt\quad (kernels, $\times 2$)\qquad
 SGEMM ($\times 2$)}};

\node[stepbox=blue!70!black, below=0.07cm of recon] (perm2)
 {\textbf{9.}\ Inverse permute: restore canonical bond ordering, per site%
 \sw{gpu\_tensor\_permute}};

\node[stepbox=teal, below=0.07cm of perm2] (deabs)
 {\textbf{10.}\ \textsc{pseudo\_sqrt} again (per site, per env bond);\quad
 deabsorb: $\psi_{v_i} \gets \mu^{-1/2}\!\cdot\!\mathbf{cv}_i$%
 \sw{symmetrize\_copy\quad SSYEVD\quad form\_sqrt\_matrices\quad
 SGEMM-batched}};

\node[stepbox=gray!55!black, below=0.07cm of deabs] (upd)
 {\textbf{11.}\ Write back:\ $\psi_{v_1},\psi_{v_2}\gets\mathbf{cv}_{1,2}$;\quad
 update bond $\chi_{\mathrm{new}}$;\quad
 $\mu \gets \mathrm{diag}(\lambda)$\ (Vidal gauge)%
 \sw{memcpy D2D ($\times 2$)\qquad memcpy H2D (diagonal message)}};

\node[outerbox=blue, below=0.35cm of upd] (rz2)
 {$R_z(h\,\delta t)$: apply unary phase gate to all sites%
 \sw{apply\_unary\_gate}};

% ── Phase 2: BP ───────────────────────────────────────────────────────────────
\node[outerbox=green!60!black, below=0.40cm of rz2] (bp)
 {Belief propagation: iterate message updates until
 $\max_e\delta_e < 10^{-5}$ or 100 sweeps%
 \sw{bp\_update}};

% ── Phase 3: Measure ──────────────────────────────────────────────────────────
\node[outerbox=purple!70!black, below=0.30cm of bp] (meas)
 {Measure $C(t)$: Bethe partition function in log-space with explicit sign tracking%
 \sw{autocorrelation\_with\_bp}};

% ── Arrows ────────────────────────────────────────────────────────────────────
\foreach \from/\to in {
 init/forloop, forloop/rz1, rz1/colorloop, colorloop/psqrt,
 psqrt/absorb, absorb/perm1, perm1/qr, qr/rr4, rr4/gate,
 gate/svd, svd/recon, recon/perm2, perm2/deabs, deabs/upd,
 upd/rz2, rz2/bp, bp/meas}
 \draw[arr] (\from) -- (\to);

% ── Group box: gate pipeline ──────────────────────────────────────────────────
\begin{pgfonlayer}{background}
 \node[groupfit, fit=(colorloop)(psqrt)(absorb)(perm1)(qr)(rr4)
 (gate)(svd)(recon)(perm2)(deabs)(upd),
 label={[font=\scriptsize\itshape,gray,rotate=90,anchor=south]left:
 \texttt{apply\_binary\_gate}}] (gatebox) {};
\end{pgfonlayer}

\end{tikzpicture}%
}
\caption{%
 \textbf{CppSim computational pipeline for one time step.}
 Each box names the mathematical operation and the GPU primitive that executes it.
 The inner group (dashed) is \texttt{apply\_binary\_gate}, called once per bond per
 color class ($\leq\!24$ times on a $4\!\times\!4$ grid).
 Color code: \textcolor{teal!70!black}{teal} = environment conditioning;
 \textcolor{blue!70!black}{blue} = linear algebra (QR/permute);
 \textcolor{orange!80!black}{orange} = gate contraction and SVD;
 \textcolor{red!60!black}{red} = tensor reconstruction;
 \textcolor{green!50!black}{green} = belief propagation;
 \textcolor{purple!70!black}{purple} = observable measurement.
 The only CPU computation in the hot path is the SVD truncation threshold (step~7),
 requiring a $D\chi$-float D2H transfer; all other steps execute on-device.
}
\label{fig:pipeline}
\end{figure}
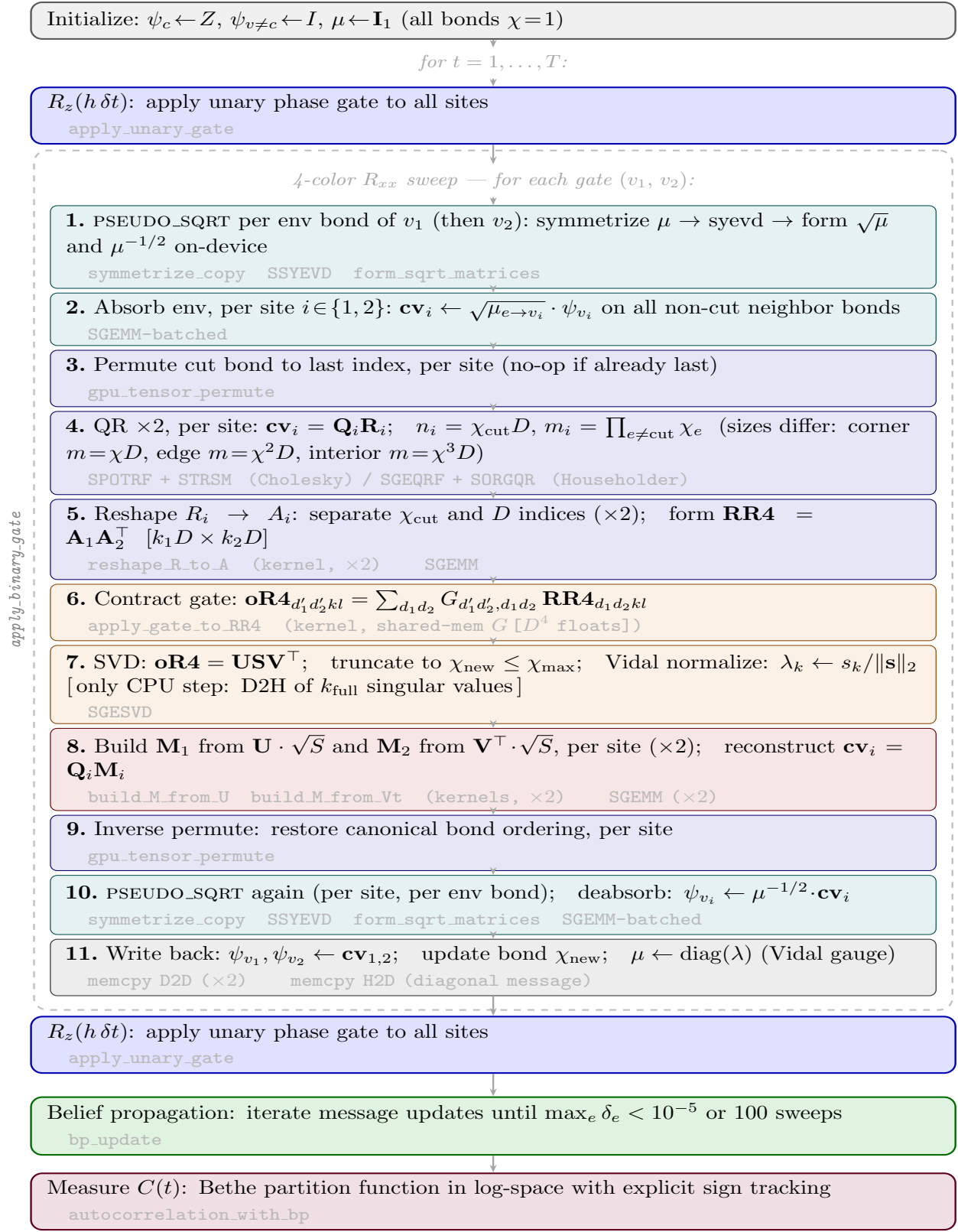
% ── End pipeline figure ───────────────────────────────────────────────────────

\paragraph{Analytic Memory Model.}
\label{sec:memory_model}

The total GPU memory requirement for a $k \times l$ grid at bond dimension $\chi$ is:
\begin{align}
 \text{State:}\quad
 M_\psi &= \bigl(4\chi^2 + 2(k+l-4)\chi^3 + (k-2)(l-2)\chi^4\bigr)
 \cdot D \cdot \texttt{sizeof(float)}, \label{eq:mem_psi} \\
 M_\mu &= 2\bigl(k(l-1)+(k-1)l\bigr)\,\chi^2 \cdot \texttt{sizeof(float)},
 \label{eq:mem_msg} \\
 \text{Workspace:}\quad
 M_W &= \bigl(4\,\chi^3 \cdot D\chi + 9\,(D\chi)^2 + 8\,\chi^2\bigr)
 \cdot \texttt{sizeof(float)}.
 \label{eq:mem_ws}
\end{align}
The dominant terms are $M_\psi \sim (k-2)(l-2)\chi^4 \cdot D$ (interior psi tensors)
and $M_W \sim 4D\chi^4$ (large workspace buffers). For a $4\times4$ grid both terms
are comparable at $\chi \geq 50$.

% ══════════════════════════════════════════════════════════════════════════════
% §6 Numerical Experiments — new content in separate file
% ══════════════════════════════════════════════════════════════════════════════
\section{Numerical Experiments}
\label{sec:experiments}
% ══════════════════════════════════════════════════════════════════════════════

All experiments use $h = J = -1.0$, $\delta t = 0.1$,
$\epsilon_{\mathrm{cut}} = 10^{-11}$, and float32 precision,
unless otherwise noted.
Although $\epsilon_{\mathrm{cut}} = 10^{-11}$ lies below the float32 noise
floor ($\varepsilon_{\mathrm{mach}} \approx 10^{-7}$), it is compatible with
float32 arithmetic because LAPACK \texttt{SGESVD} returns singular values
as float32 scalars representable down to $\sim10^{-38}$, and the threshold is
applied as a relative comparison $s_i \geq \epsilon_{\mathrm{cut}}\,s_0$.
In practice, all physically meaningful singular values lie well above
$10^{-7}\,s_0$; the cutoff acts only to discard subnormal numerical zeros
produced by rounding, leaving $\chimax$ as the active truncation constraint
in every experiment.
The center observable is $Z_c$ at site $(\lfloor(N_x-1)/2\rfloor,\lfloor(N_y-1)/2\rfloor)$,
the geometric center of the grid.

The primary observable throughout is the \emph{spin autocorrelation function}
\begin{equation}
 C(t) = \frac{1}{2^N}\,\mathrm{Tr}\bigl[Z_c(t)\,Z_c\bigr],
 \qquad
 Z_c(t) = e^{i\mathcal{H}t}\,Z_c\,e^{-i\mathcal{H}t},
 \label{eq:ct_def}
\end{equation}
where $N$ is the number of sites and $Z_c$ is the Pauli-$Z$ operator at the
lattice center.
$C(t)$ measures the overlap between the Heisenberg-evolved operator $Z_c(t)$
and its initial form $Z_c$: it equals 1 at $t=0$ by normalization, and
decreases toward 0 as $Z_c(t)$ scrambles into many-body Pauli strings
that are orthogonal to $Z_c$.
This decay is purely \emph{unitary} --- there is no dissipation or energy loss;
$C(t) \to 0$ means the operator has spread across the entire lattice,
not that it has decayed.

\subsection{Correctness Validation}
\label{sec:correctness}

Validation proceeds on two levels.
The first establishes \emph{implementation correctness} by comparison with an
existing reference implementation.
The second establishes \emph{physical fidelity} through consistency with
predictions derived independently of any code.
This second level is not a cross-check of an existing answer: it is the
instrument through which the physical behavior of the system was understood,
and through which the results of \S\ref{sec:bp_convergence} --- including the
connection between operator-placement symmetry and BP fixed-point multiplicity
--- were discovered.
The distinction matters: where the reference begins from adapted library codes,
the implementation reported here codes the physics directly --- each component
derived from first principles and validated against the physical predictions it
is designed to satisfy.

\medskip
\noindent\textbf{Implementation correctness} is established against two external references:
\begin{enumerate}
 \item \textbf{Julia reference}: $\Ct$ for $4\times4$ at $\chi=50$ matches the
 Julia float64 implementation to within float32 rounding
 ($<10^{-6}$ absolute error) for $t \leq 1.6$.
 \item \textbf{Chi consistency}: for $t \leq 1.4$, $\Ct$ agrees across
 $\chi \in \{40, 50, 60, 80, 110\}$ to 5 significant figures
 (Figures~\ref{fig:ct_4x4} and~\ref{fig:ct_5x5}), confirming that truncation error is negligible
 for early-to-mid time evolution at these bond dimensions.
\end{enumerate}

\medskip
\noindent\textbf{Physical fidelity} is established by consistency with predictions
derived independently of any code.
The simulation is correct not only when it matches a reference, but when it
reproduces the qualitative and quantitative behavior that the physics demands.
\begin{enumerate}
 \item \textbf{Operator spreading and lightcone}: under unitary evolution,
 $Z_c(t)$ must spread outward from the center site.
 The non-identity Pauli weight $n(r,t)$ should grow from a single pixel
 at $t=0$ to a spatially extended pattern, bounded by the Lieb--Robinson
 cone~\citep{liebrobinson}.
 The simulation reproduces this: $n(r,t)=0$ at all non-center sites at
 $t=0$, and the spreading front reaches the lattice boundary at
 $n^*$ layers (Table~\ref{tab:lightcone}), consistent with the predicted
 lightcone velocity.
 \item \textbf{Early-time universality}: for $t \leq 1.4$ the operator front
 has not yet reached any lattice boundary regardless of grid size, so
 $\Ct$ must be independent of both $\chi$ and grid size.
 All eight $\chi$ values and all eight grid sizes agree to 5 significant
 figures in this regime (Figures~\ref{fig:ct_4x4} and~\ref{fig:ct_5x5}; Table~\ref{tab:ct_grids}).
 \item \textbf{$C_4$ symmetry of $n(r,t)$}: placing $Z_c$ at the geometric
 center of an odd grid imposes $C_4$ symmetry on the operator lightcone.
 The simulation reproduces exact four-fold symmetry in $n(r,t)$ at every
 Trotter layer, to 6 significant figures in all corner weights --- a
 sensitive check that the center formula, gate application, and BP update
 are all implemented correctly.
 \item \textbf{Scrambling to zero}: at long times, $Z_c(t)$ spreads into
 many-body Pauli strings orthogonal to $Z_c$, so $\Ct \to 0$.
 The simulation reaches $\Ct < 10^{-3}$ by $t \approx 4$ on all grids
 large enough to avoid finite-size reflections, and $n(r,t)$ saturates
 uniformly across all sites, both consistent with full operator scrambling.
\end{enumerate}
These physical consistency checks validate aspects of the simulation that
numerical comparisons alone cannot reach: a wrong center formula or a broken
symmetry in the gate implementation would violate items~3 and~1 respectively,
even if $\Ct$ happened to match a reference at early times.

\paragraph{Runtime diagnostic measures.}
Agreement with the Julia reference and chi consistency together confirm that
$\Ct$ is correct when the simulation runs cleanly.
However, $\Ct$ alone cannot distinguish a correct answer from a physically
plausible artifact: a BP trap or an over-saturated bond dimension can produce
a smooth, monotone $\Ct$ trajectory that deviates from the true value by
$5$--$10\%$ without any obvious signature in $\Ct$ itself.
To detect such failures before they corrupt the result, \textsc{CppSim} reports
four auxiliary diagnostics per Trotter layer, listed in
Table~\ref{tab:diagnostics}.

\begin{table}[ht]
\small\centering
\caption{Runtime diagnostic measures reported per Trotter layer,
 spanning the full failure chain: truncation $\to$ BP $\to$
 Bethe approximation $\to$ bond capacity.}
\label{tab:diagnostics}
\begin{tabular}{lll}
\toprule
Quantity & Definition & Failure signal \\
\midrule
\texttt{gate\_err} &
 $\sum_{\text{bonds}} \sum_{i>\chi} s_i^2 \,/\, \sum_i s_i^2$ &
 $> 10^{-1}$: $\Ct$ qualitative only \\
Bethe norm &
 $\exp\!\bigl(\sum_v \log f_v - \sum_e \log Z_e\bigr)$ &
 $\geq 2.0$: \texttt{norm\_exploded} flag \\
BP sweeps\,/\,$\delta$ &
 iterations to $\max_e|\Delta\mu_e| < \epsilon_{\mathrm{tol}}$ &
 sweeps\,$\geq 30$ \& $\delta < \epsilon_{\mathrm{tol}}$: trapped layer (\S\ref{sec:bp_convergence}) \\
$S_e / \log\chi$ &
 $-\sum_k \lambda_k^2 \log\lambda_k^2 \,/\, \log\chi$ &
 $\geq 0.5$: increase $\chi$, not $\alpha$ (BP damping, \S\ref{sec:bp_convergence}) \\
\bottomrule
\end{tabular}
\end{table}

The four diagnostics address distinct points in the simulation pipeline.
\texttt{gate\_err} captures SVD truncation at the gate level: it is the
fraction of singular value weight discarded by the $\chi$ cutoff, summed
over all bonds in the layer.
A value below $10^{-2}$ means truncation is not the limiting factor;
values above $10^{-1}$ (seen at $\chi=20$ on large grids) signal that
$\Ct$ is qualitative rather than quantitative.
The Bethe norm measures the self-consistency of the BP environment:
it equals 1 exactly on a tree and deviates when the Bethe approximation
breaks down on the loopy graph --- a graph containing cycles, where BP is an
approximation rather than exact~\citep{yedidia2005constructing}.
Values above 2.0 trigger the \texttt{norm\_exploded} flag and typically
correlate with incorrect $\Ct$; the norm can recover spontaneously in the
next layer as BP re-converges, so a transient spike does not necessarily
invalidate the run.
The BP sweep count and final $\delta$ jointly identify convergence quality:
the combination (sweeps $\geq 30$, $\delta < \epsilon_{\mathrm{tol}}$) is the
early-warning signature of BP converging to a wrong fixed point --- a
\emph{trapped layer}, analyzed in detail in \S\ref{sec:bp_convergence}.
Finally, $S_e / \log\chi$ is the per-bond entanglement saturation fraction,
where $S_e = -\sum_k \lambda_k^2 \log\lambda_k^2$ is the operator-space
entanglement entropy (OSEE) of the Vidal singular values on bond $e$.
When $S_e / \log\chi \geq 0.5$, the bond is more than half-saturated and
the variational manifold at bond dimension $\chi$ is approaching its capacity;
increasing the BP damping factor $\alpha$ cannot compensate, and increasing $\chi$ is the only remedy
(\S\ref{sec:bp_convergence}, capacity boundary).

% ── Chi convergence figures and trap table ────────────────────────────────────
\subsection{$\chi$-Convergence of $C(t)$}
\label{sec:chi_convergence}

\begin{figure}[ht]
\centering
\includegraphics[width=\linewidth]{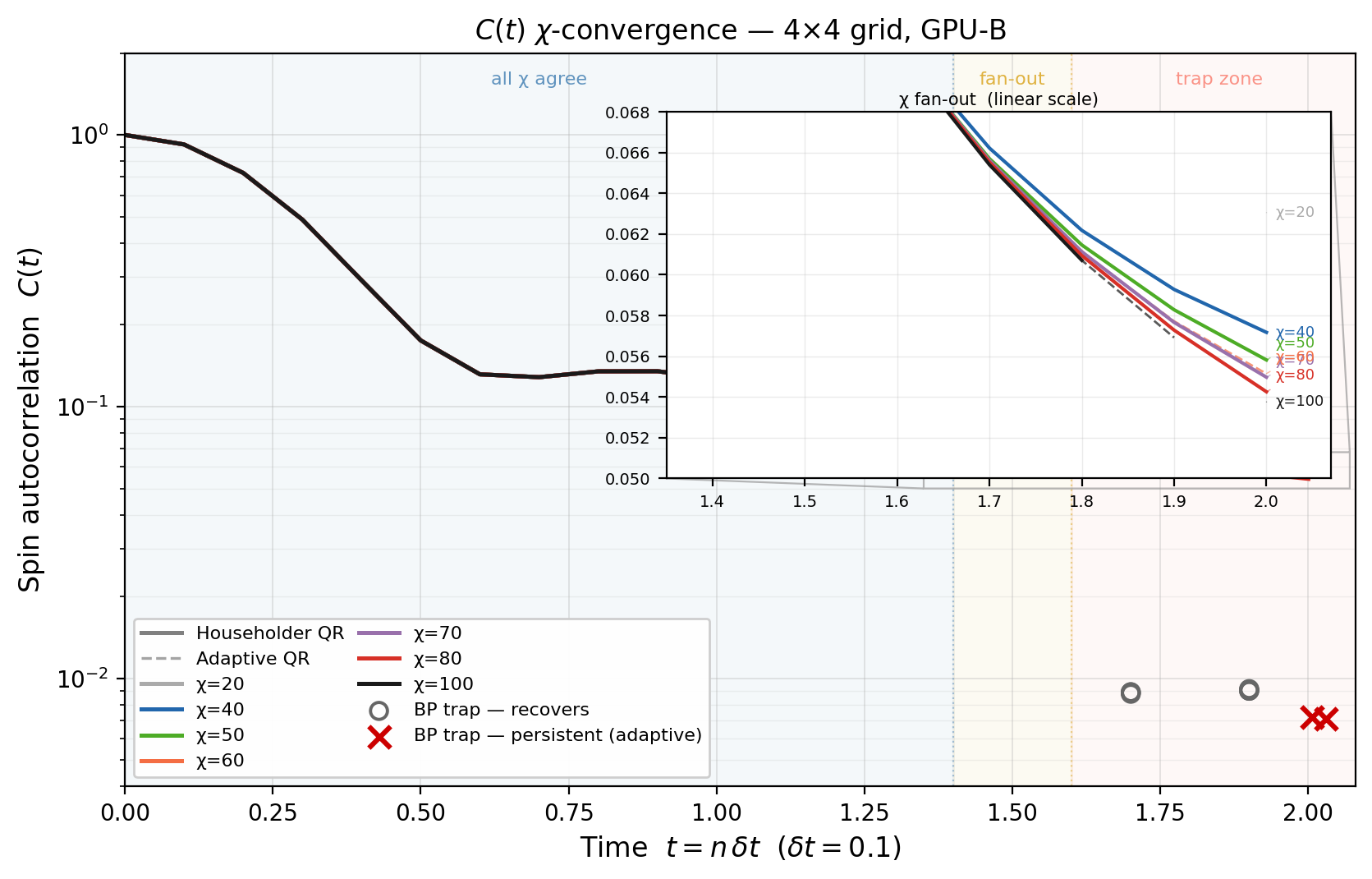}
\caption{%
 Spin autocorrelation $\Ct$ on a $4\times4$ grid (GPU-B, log scale) for
 $\chi \in \{20, 40, 50, 60, 70, 80, 100\}$.
 Householder QR: solid lines. Adaptive QR: dashed lines (same color per $\chi$).
 Three dynamical regimes are shaded:
 \emph{all $\chi$ agree} ($t \leq 1.4$, both methods overlap to 5 significant
 figures);
 \emph{fan-out} ($1.4 < t \leq 1.6$, $\chi$-dependent truncation error becomes
 visible, detail in inset);
 \emph{trap zone} ($t > 1.6$, BP metastable traps may occur).
 Open circles: BP traps that recover at the next layer (Householder QR).
 Red crosses: persistent traps that do not recover by $t=2.0$ (Adaptive QR,
 $\chi=70$ and $\chi=100$).
 Inset (linear scale): late-time $\chi$ fan-out with per-$\chi$ labels;
 Householder convergence is monotone from $\chi=40$ to $\chi=100$,
 extrapolating to $\Ct(2.0) \approx 0.052$--$0.053$.}
\label{fig:ct_4x4}
\end{figure}

\begin{figure}[ht]
\centering
\includegraphics[width=\linewidth]{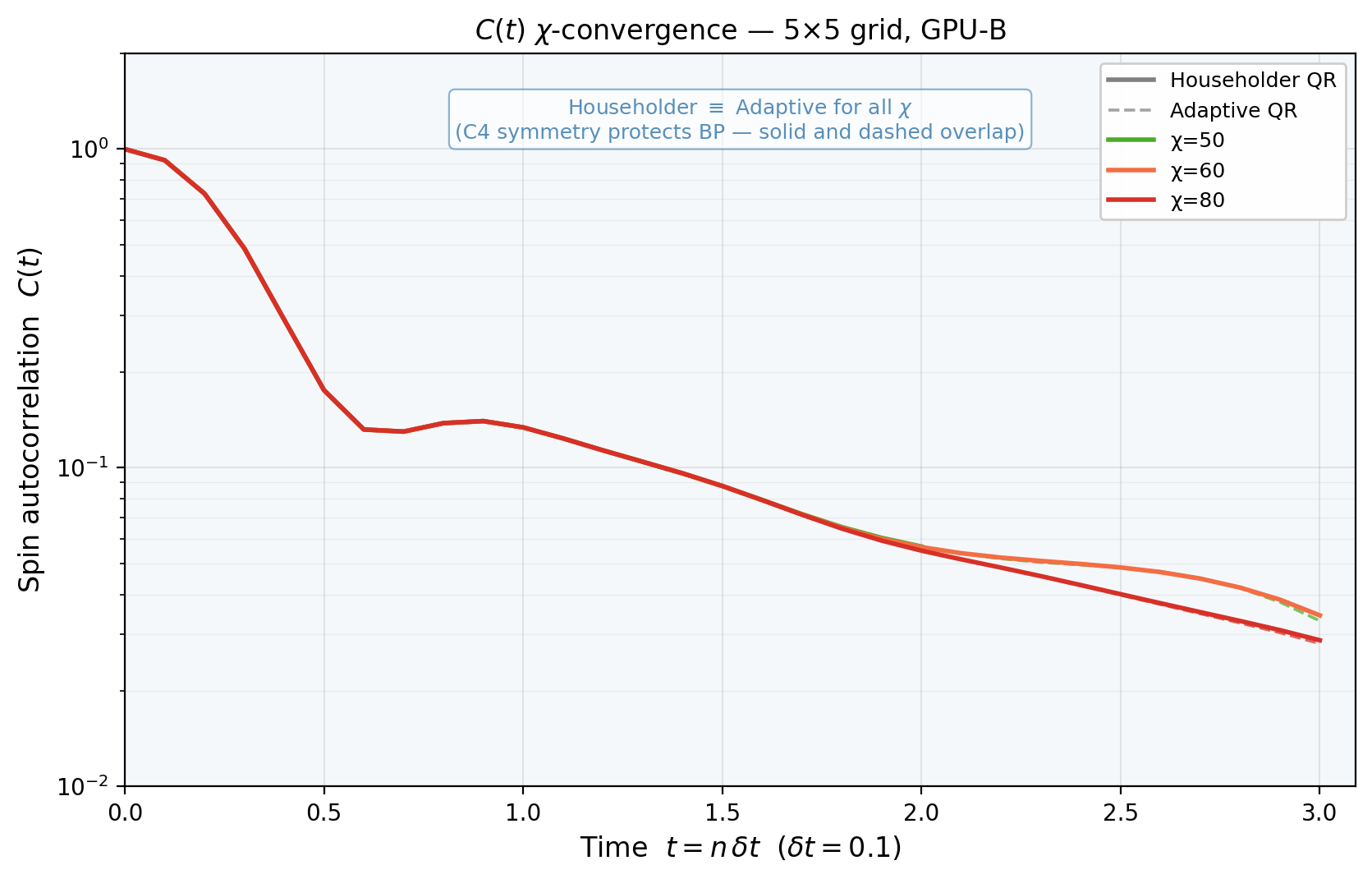}
\caption{%
 Spin autocorrelation $\Ct$ on a $5\times5$ grid (GPU-B, log scale) for
 $\chi \in \{50, 60, 80\}$, 30 Trotter layers.
 Householder QR: solid lines. Adaptive QR: dashed lines.
 The two methods are indistinguishable at every layer: the C4 lattice symmetry
 of the centered initial condition prevents BP from finding metastable fixed
 points regardless of gauge, so Adaptive QR produces identical physics to
 Householder QR on this geometry.
 The $\chi$-convergence is slower than on $4\times4$: curves for $\chi=50$,
 $60$, and $80$ remain visibly separated at $t=3.0$, consistent with larger
 entanglement growth on the bigger grid.}
\label{fig:ct_5x5}
\end{figure}

Throughout this section, GPU-B denotes the 32\,GB, 1.2\,TB/s configuration
and GPU-A the 32\,GB, 1.0\,TB/s configuration
(Section~\ref{sec:portability}).
Figures~\ref{fig:ct_4x4} and~\ref{fig:ct_5x5} show $\Ct$ for $\chi \in
\{20,\ldots,100\}$ on $4\times4$ and $5\times5$ grids (GPU-B), comparing
Householder QR (solid lines) against Adaptive QR (dashed lines).

The $4\times4$ results reveal three regimes:

\paragraph{Early time ($t \leq 1.4$): universal agreement.}
All $\chi$ values and both QR modes agree to 5 significant figures,
confirming that bond dimension plays no role until entanglement saturates the
smallest bond.

\paragraph{Monotone convergence for clean runs ($t > 1.4$).}
The Householder convergence curve is strictly decreasing in $\chi$ at every
late layer.
At $t=2.0$: $\Ct(\chi=40) > \cdots > \Ct(\chi=100)$, with values
$0.0572 > 0.0558 > 0.0552 > 0.0550 > 0.0543$,
extrapolating to $\approx 0.052$--$0.053$.
We identify $\chi^* \approx 60$ as the point of diminishing returns:
below it each increment in $\chi$ buys meaningful accuracy; above it the gain
is less than $0.2\%$ per step while cost grows as $\chi^4$.

\paragraph{Trap zone ($t > 1.6$).}
BP metastable traps appear stochastically.
Householder traps are transient --- they recover at the next layer and do not
contaminate the $\chi$-convergence curve.
Adaptive traps at $\chi=70$ and $\chi=100$ are persistent at $t=2.0$
(final layer), producing $\Ct \approx 0.007$ instead of $\approx 0.055$.

Table~\ref{tab:trap_inventory} catalogues all observed BP traps across both
grids, QR modes, and GPU architectures.

\begin{table}[ht]
\centering
\caption{BP trap inventory: layer indices where $\Ct$ collapsed to
 $\approx 0.007$--$0.010$, and whether the trap recovered at the next layer.
 \emph{Recovers}: correct value restored at layer $+1$.
 \emph{Persistent}: final layer still trapped, no further chance to recover.
 GPU-A runs (both QR modes) show more trapping than GPU-B at the same $\chi$ due to
 floating-point non-determinism (matrix multiply-accumulate units (GPU-B) versus scalar vector-ALU (GPU-A)).
 \emph{n/a}: that $\chi$ was not run on this architecture/mode combination.
 No traps occur on the $5\times5$ grid in either method.}
\label{tab:trap_inventory}
\small
\begin{tabular}{cllll}
\toprule
$\chi$ & GPU-B Adaptive & GPU-B Householder & GPU-A Adaptive & GPU-A Householder \\
\midrule
20 & --- & L17, L19 (recover) & L17 (recovers) & L17, L19 (recover) \\
40 & --- & --- & n/a & --- \\
50 & --- & --- & L20 (\textbf{persistent}) & L19 (\textbf{persistent}) \\
60 & --- & L19 (recovers) & n/a & L12--L20 (\textbf{persistent}) \\
70 & L20 (\textbf{persistent}) & L17 spike (recovers) & n/a & L6, L16, L19, L20 (\textbf{persistent}) \\
80 & --- & --- & L16, L18, L20 (\textbf{persistent}) & --- \\
90 & n/a & L19 (recovers) & n/a & n/a \\
100 & L20 (\textbf{persistent}) & L19 (recovers) & n/a & n/a \\
\midrule
\multicolumn{5}{l}{\emph{$5\times5$, all $\chi$}: no traps in either method (C4 symmetry protects BP).} \\
\bottomrule
\end{tabular}
\end{table}

% ── BP fixed-point section ────────────────────────────────────────────────────
\subsection{BP Fixed-Point Multiplicity, Root Cause, and Trap Remediation}
\label{sec:bp_convergence}

Figure~\ref{fig:ct_4x4} shows isolated anomalous layers --- marked with open circles
and red crosses --- where $\Ct$ collapses to $\approx 0.007$--$0.009$ while all
neighboring layers follow the smooth converging sequence.
These \emph{trapped layers} are symptoms of a deeper structural property of the
BP fixed-point landscape.
Before presenting the experimental data, we define the two distinct ways BP can
fail, since both appear in the results and must not be confused.

\paragraph{Two failure modes.}
\emph{Type~1 --- metastable trap (trapped layer)}: BP converges in a finite
number of sweeps ($\delta < \epsilon_{\mathrm{tol}} = 10^{-5}$), yet $\Ct$
collapses to $\approx 0.007$--$0.010$, inconsistent with all neighboring layers.
In optimization terms, the iterative message-passing dynamics have settled into
a local basin of the Bethe free energy landscape --- a local fixed point that is
stable under further BP iteration but is not the global optimum.
The word \emph{trap} captures this precisely: the algorithm is not diverging;
no amount of additional sweeps at the same $\alpha$ will escape the basin.
The sweep count at a trapped layer is $2$--$3\times$ the typical value
(e.g., 49 sweeps at $\chi=110$, $t=1.7$ versus 12--18 on neighboring layers),
providing an early-warning signal detectable at runtime.

\emph{Type~2 --- non-convergence}: BP exhausts the 100-sweep budget with
$\delta \gg \epsilon_{\mathrm{tol}}$.
The resulting $\Ct$ is unreliable at that layer, but the next Trotter layer
typically recovers; the event is self-announcing (sweep count $= 100$, large
$\delta$).
Type~2 does not imply a wrong answer: GPU-A $\chi=50$, $t=1.9$ fails to
converge (100 sweeps, $\delta = 1.80\times10^{-5}$) yet produces $\Ct = 0.060$,
agreeing with the GPU-B clean run to within 3\%.

The sweep count alone does not distinguish the types; the final $\delta$ is
essential.
A runtime diagnostic should flag any layer where (sweeps $\geq 30$) \emph{and}
($\delta < \epsilon_{\mathrm{tol}}$) as a candidate Type~1 trapped layer,
warranting a retry.
With this vocabulary in place, we now identify the root cause.

\paragraph{Root cause: operator-placement symmetry governs fixed-point multiplicity.}
When $Z_c$ is placed at the geometric center of an odd grid, the operator
lightcone has exact $C_4$ symmetry from the first Trotter layer: all four
quadrants of the lattice are equivalent, so symmetry-equivalent bonds carry
identical Schmidt spectra and BP messages at the correct fixed point must also
be equal.
This constraint \emph{collapses} the effective degrees of freedom that BP must
resolve, reducing the number of competing fixed points in the Bethe free energy
landscape.
Breaking this symmetry --- by displacing $Z_c$ even one site, or by using a
grid geometry where no site achieves true $C_4$ symmetry --- lifts the
constraint, populating the message manifold with additional local minima and
making traps more likely.

We test this hypothesis with a $5\times5$ grid, where the geometric center
$(2,2)$ achieves exact $C_4$ symmetry (all four corners at Manhattan distance
$d_{\max}=4$), unlike the $4\times4$ grid where no center site achieves this.
The experiment runs $\chi \in \{20, 40, 60, 80, 100\}$ on the $5\times5$
symmetric grid and compares to the $4\times4$ results (Table~\ref{tab:trap_inventory}).

The result is \emph{zero} Type~1 trapped layers on the symmetric $5\times5$ grid
across all $\chi$ values and all layers (Figure~\ref{fig:ct_5x5}),
with final values $\Ct(\chi=20)=0.000716$, $\Ct(\chi=40)=0.000907$,
$\Ct(\chi=60)=0.001073$, $\Ct(\chi=80)=0.028089$, $\Ct(\chi=100)=0.042065$.
The $5\times5$ grid has strictly more loops ($40$ bonds versus $24$ for $4\times4$),
and the data in Table~\ref{tab:trap_inventory} already show
that larger grids and higher bond dimensions increase the complexity of the BP
update and the density of the entanglement spectrum.
One might therefore expect more trapped layers on the $5\times5$ grid, not
fewer.
Symmetry proves to be the dominant factor: on the symmetric grid, BP sweeps
produce only Type~2 non-convergence events at high saturation
($S_e/\log\chi \gtrsim 0.65$, late layers, $\Ct \approx 0$) --- self-announcing
events that do not corrupt $\Ct$ --- but never settle into a wrong fixed point.
At $\chi=80$, the trap that struck at $t=1.7$ on the $4\times4$ grid
($\Ct=0.0088$) is entirely absent: the $5\times5$ symmetric run at $t=1.7$
gives $\Ct=0.071$ with $11$ sweeps, physically consistent with all neighbors.

The mechanism is as follows.
On a $C_4$-symmetric lattice with a $C_4$-symmetric initial operator, the
correct BP fixed point respects the symmetry: messages on related bonds are
equal.
This symmetry-constrained fixed point is unique in the symmetric subspace.
When the operator breaks $C_4$ symmetry, the full (unsymmetrized) message
manifold is active, and multiple local minima can coexist at the same $(\chi,
t)$.
The floating-point arithmetic path --- determined by GPU reduction order ---
then selects among them, producing the architecture-dependent, non-monotone
trap patterns observed on the $4\times4$ grid.

\paragraph{Further evidence: architecture and update-rule experiments.}
Two additional controlled experiments corroborate the fixed-point multiplicity
picture on the $4\times4$ grid.

\emph{Cross-architecture experiment.}
The trap patterns in Table~\ref{tab:trap_inventory} are \emph{complementary} across
architectures: GPU-B traps at $(\chi=80,\,t=1.7)$ and $(\chi=70,\,t=2.0)$
while GPU-A is clean at those points; conversely, GPU-A traps at
$(\chi=50,\,t=2.0)$ and $(\chi=80,\,t=2.0)$ while GPU-B is clean.
Both GPUs execute the same algorithm in float32, but matrix multiply-accumulate units (GPU-B) and
scalar vector-ALU (GPU-A) reduction trees differ in rounding order,
perturbing the message trajectory enough to select a different basin.
When both are un-trapped, $\Ct$ agrees to within $1$--$2\times10^{-3}$.

\emph{Update-rule experiment.}
Fixing $\chi=80$ on GPU-B, we compare undamped BP ($\alpha=1$) and damped BP
with $\alpha \in \{0.50,\, 0.80\}$.
Damping introduces a momentum term into the message update:
\begin{equation}
 \mu_e^{(k+1)}
 \;\leftarrow\;
 (1-\alpha)\,\mu_e^{(k)} \;+\; \alpha\,\hat{\mu}_e^{(k+1)},
 \label{eq:damping}
\end{equation}
where $\hat{\mu}_e^{(k+1)}$ is the undamped update.
Spectrally, damping maps an eigenvalue $\lambda$ of the BP Jacobian to
$(1-\alpha) + \alpha\lambda$, compressing the spectrum toward 1 and
suppressing oscillatory modes that otherwise overshoot the fixed point.
The result at $(\chi=80, t=1.7)$: $\alpha=0.50$ avoids the trap entirely
($\Ct = 0.066$), while $\alpha=0.80$ and undamped both find the wrong basin
($\Ct = 0.009$, identical to six decimal places).
At $(\chi=80, t=1.9)$: the roles reverse --- $\alpha=0.80$ finds the correct
fixed point ($\Ct = 0.057$) while $\alpha=0.50$ does not converge.
Both experiments are consistent with the symmetry picture: the $4\times4$ grid
has multiple fixed points, and any perturbation to the message trajectory
(architecture, damping) selects among them.

\paragraph{Trap remediation.}
A key practical conclusion is that traps are \emph{not fatal}: the correct
$\Ct$ can be recovered by retrying the BP step with a different $\alpha$.
This is confirmed by the cross-$\alpha$ experiment: GPU-B damped $\alpha=0.50$
at $(\chi=80, t=1.7)$ recovers $\Ct = 0.066$, matching GPU-A undamped
($\Ct = 0.066$) to within $1.5\times10^{-3}$ --- two different hardware paths
arriving at the same physical fixed point.
The most robust long-term remedy is to use a symmetric initial condition
(odd grid with $Z_c$ at the geometric center), which eliminates Type~1 traps
entirely.
For asymmetric configurations, the remediation procedure is:
\begin{enumerate}
 \item After each BP run, check whether (sweeps $\geq 30$) and
 (delta $< \epsilon_{\mathrm{tol}}$).
 If so, flag as candidate Type~1 trap.
 \item Verify by comparing $\Ct$ to the previous layer: a drop of $>50\%$
 with no corresponding increase in \texttt{gate\_err} is a strong signal.
 \item Retry BP with a different $\alpha$ (e.g., try $\alpha=0.50$ if
 undamped trapped; try $\alpha=0.80$ if $\alpha=0.50$ trapped).
 \item If all $\alpha$ values fail simultaneously, the bond dimension is
 insufficient: increase $\chi$ rather than tuning $\alpha$.
\end{enumerate}

\paragraph{Capacity boundary.}
A distinct third regime appears at high entanglement saturation
($S_e/\log\chi \gtrsim 0.65$, observed consistently across both grid sizes):
BP non-convergence (Type~2) becomes systematic and $\Ct$ approaches zero
regardless of $\alpha$.
At these layers the bond dimension itself is the bottleneck --- the variational
manifold is saturated and no message-passing strategy can compensate.
This boundary is cleanly separated from the fixed-point multiplicity regime:
in the multiplicity regime, at least one $\alpha$ recovers the correct $\Ct$;
at the capacity boundary, none does.
The appropriate response is to increase $\chi$, not to tune the update rule.

At $\chi=40$ and $\chi=60$, both GPU-B and GPU-A are trap-free and agree to
within $5\times10^{-4}$ ($\Ct(\chi=60, t=2.0)$: GPU-B $= 0.0551$,
GPU-A $= 0.0546$), confirming these as \emph{portable operating points}: the BP
landscape is simple enough that neither damping nor architecture choice matters.
The $5\times5$ symmetric runs confirm that portability extends beyond
$\chi=60$: on a symmetric grid, $\chi=80$ and $\chi=100$ are also trap-free,
suggesting that the portable range expands when physical symmetry is respected.

% ── Operator spreading and lightcone ─────────────────────────────────────────
\subsection{Operator Spreading and Lightcone Saturation}
\label{sec:lightcone}

The spin autocorrelation $\Ct = \mathrm{Tr}[Z_c(t)\,Z_c]/2^N$ measures how
much the evolved operator $Z_c(t)$ still overlaps its initial form $Z_c$ at
the center site.
Under unitary evolution, $\Ct \to 0$ does \emph{not} signal dissipation:
there is no energy loss, no bath.
Instead it signals \emph{operator scrambling}: $Z_c(t)$ has spread across
the lattice into many-body Pauli strings that are orthogonal to the initial
local $Z_c$.
The rate and geometry of this spreading are constrained by a Lieb--Robinson
bound~\citep{liebrobinson}: the operator front propagates at a group velocity
bounded by $v_{\mathrm{LR}} = 2|J|$ for the Heisenberg coupling
(equation~\eqref{eq:hamiltonian}), reaching the lattice corner at Manhattan
distance $d_{\mathrm{max}}$ after at least
\begin{equation}
 n^* = \frac{d_{\mathrm{max}}}{2|J|\,\delta t}
 \quad \text{Trotter layers,}
 \label{eq:nstar}
\end{equation}
where $\delta t$ is the Trotter step size.
In all experiments here $|J| = 1$ and $\delta t = 0.1$, so
$n^* = 5\,d_{\mathrm{max}}$.
For $n < n^*$, the boundary has not yet been reached and $\Ct$ probes
bulk scrambling; for $n \geq n^*$, finite-size effects enter and $\Ct$
approaches its long-time plateau.
Table~\ref{tab:lightcone} lists $n^*$ for each grid.

\begin{table}[ht]
\centering
\caption{Lightcone saturation: grid size, geometric center, maximum Manhattan
 distance $d_{\mathrm{max}}$ from center to corner, and minimum steps $n^*$
 to reach the boundary.
 Runs with $n < n^*$ observe bulk scrambling only.}
\label{tab:lightcone}
\begin{tabular}{ccccc}
\toprule
Grid & Center & $d_{\mathrm{max}}$ & $n^*$ (layers) & Layers run \\
\midrule
$3\times3$ & $(1,1)$ & 2 & 10 & 20 \\
$4\times4$ & $(1,1)$ & 4 & 20 & 20 \\
$5\times5$ & $(2,2)$ & 4 & 20 & 30 \\
$6\times6$ & $(2,2)$ & 6 & 30 & 30 \\
$7\times7$ & $(3,3)$ & 6 & 30 & 40 \\
$8\times8$ & $(3,3)$ & 8 & 40 & 40 \\
$9\times9$ & $(4,4)$ & 8 & 40 & 50 \\
$10\times10$ & $(4,4)$ & 9 & 45 & 50 \\
\bottomrule
\end{tabular}
\end{table}

\begin{figure}[ht]
\centering
\includegraphics[width=\linewidth]{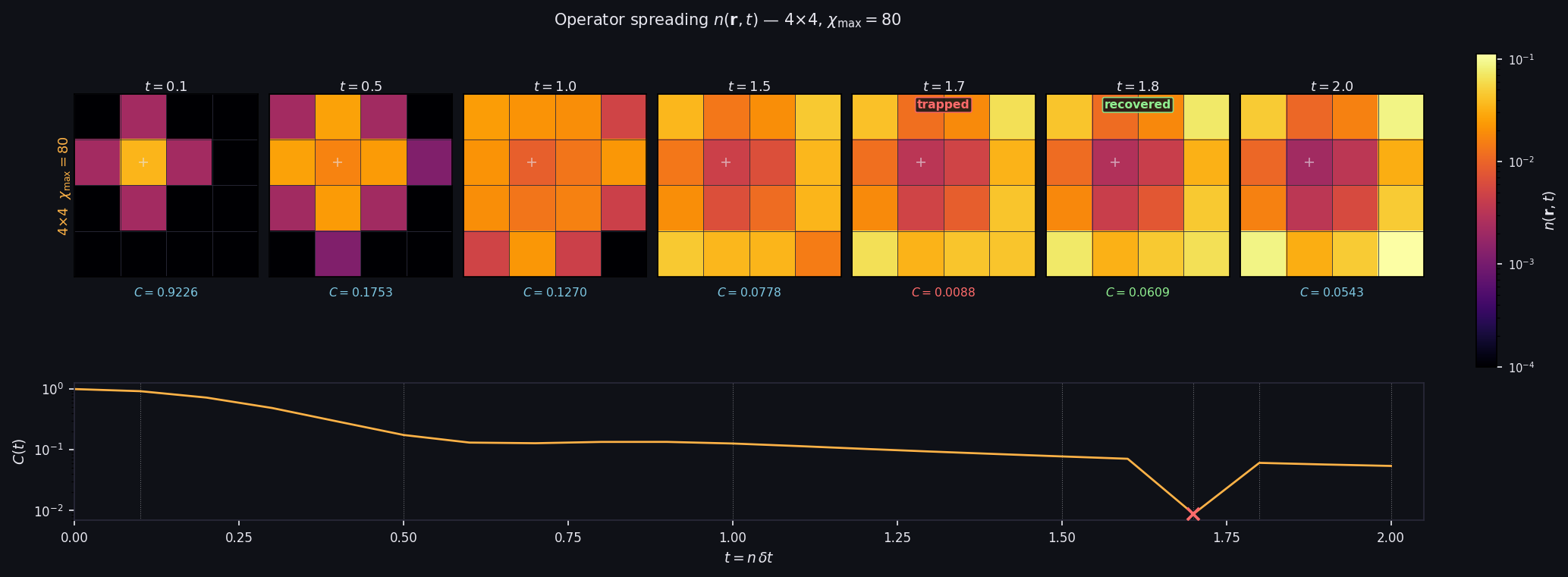}
\caption{%
 Per-site non-identity Pauli weight $n(\mathbf{r},t)$ on a $4\times4$ grid,
 $\chi_{\max}=80$ (GPU-B), seven Trotter snapshots.
 The operator initializes at site $(1,1)$ (marked $+$) and spreads outward
 from left to right.
 The panel at $t=1.7$ carries a BP metastable trap (``trapped'' banner):
 $C(t)$ collapses to $0.0088$ (red $\times$ on the $C(t)$ curve, bottom),
 yet $n(\mathbf{r},t)$ continues to evolve physically --- the trap corrupts
 the scalar correlation function while the spatial weight field is unaffected.
 The run recovers at $t=1.8$ ($C(t)$ returns to the converging sequence).
 Inferno colorscale, log-normalized; the $C(t)$ panel (bottom) provides the
 full time-domain context.
}
\label{fig:lightcone_trap}
\end{figure}

\begin{figure}[ht]
\centering
\includegraphics[width=\linewidth]{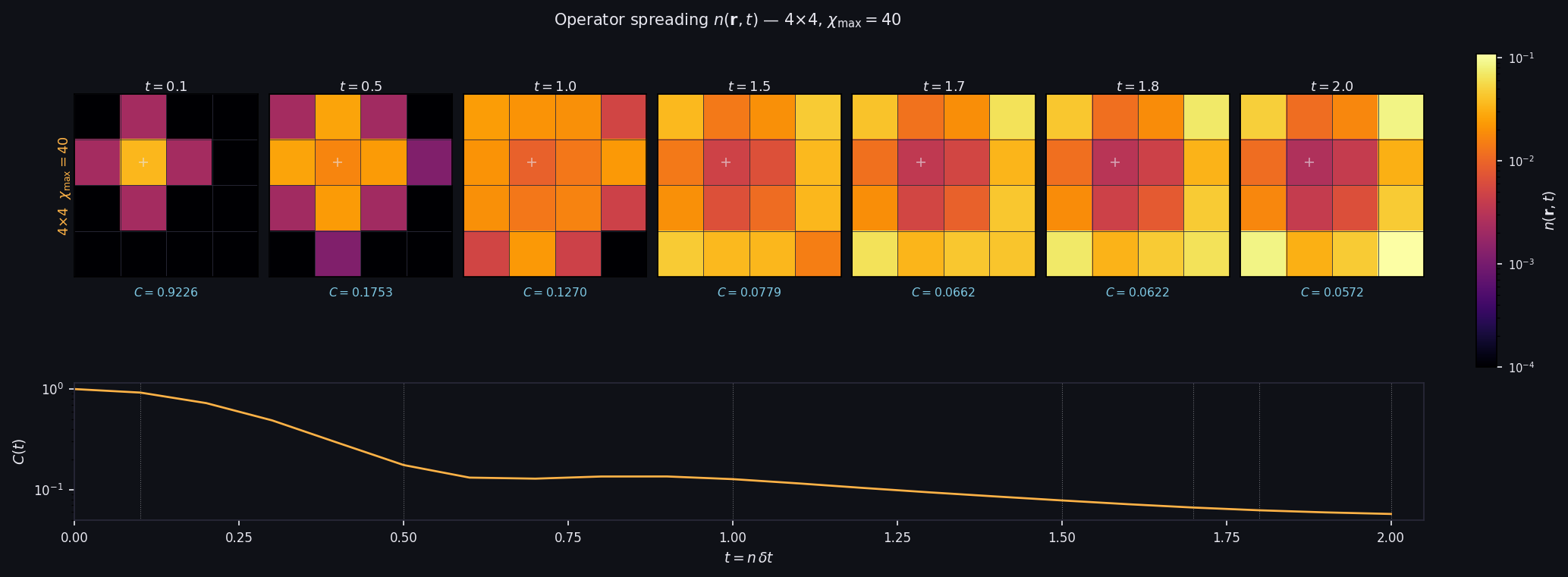}
\caption{%
 Same layout as Figure~\ref{fig:lightcone_trap} for the clean baseline run
 at $\chi_{\max}=40$ (no BP trap).
 $C(t)$ decays monotonically from $1$ to $\approx 0.057$ at $t=2.0$.
 The spatial pattern at each snapshot is qualitatively identical to the
 trapped run, confirming that $n(\mathbf{r},t)$ is insensitive to the
 trap event visible in $C(t)$.
}
\label{fig:lightcone_clean}
\end{figure}

\begin{figure}[ht]
\centering
\includegraphics[width=\linewidth]{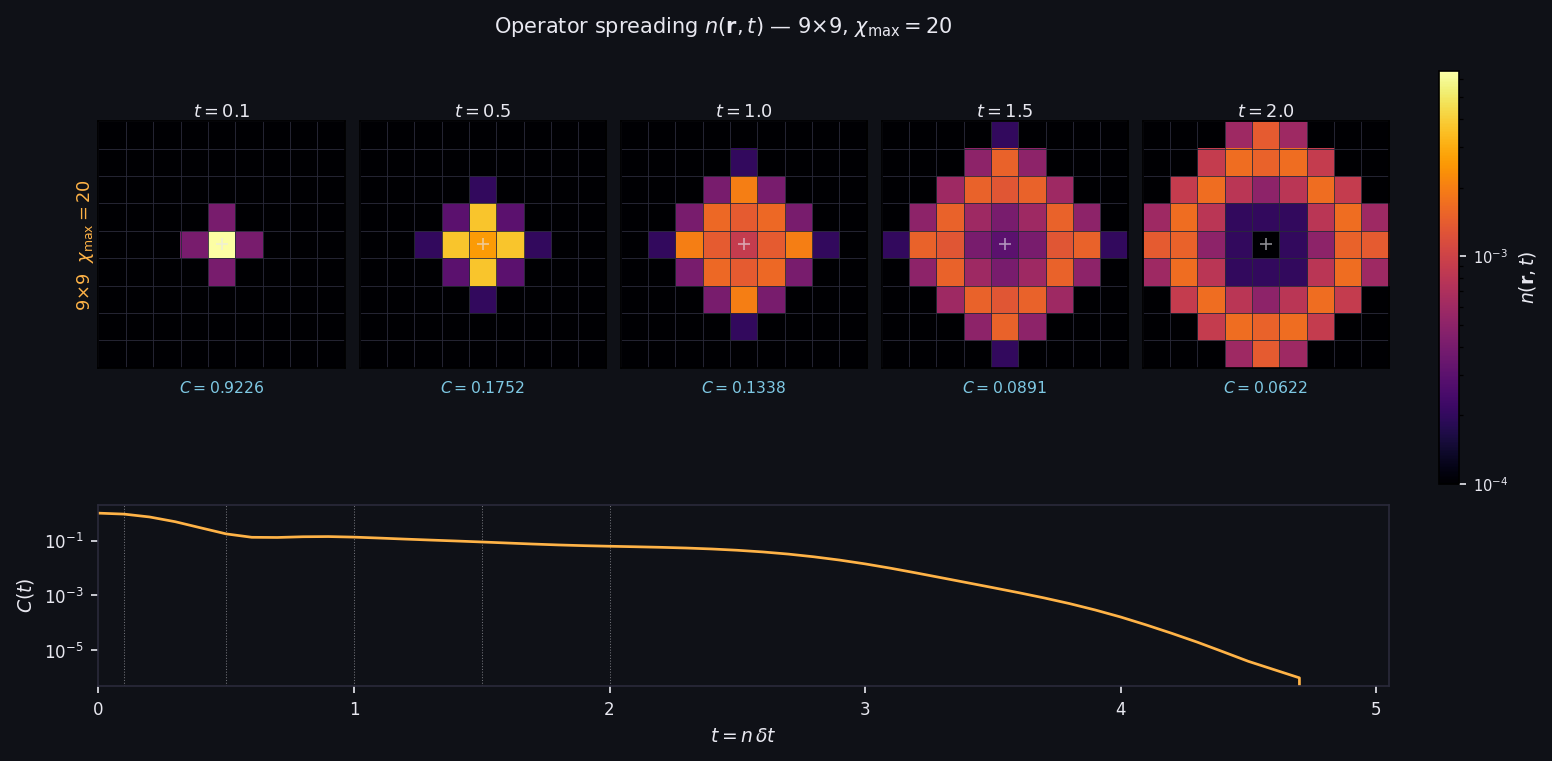}
\caption{%
 $n(\mathbf{r},t)$ on a $9\times9$ grid, $\chi_{\max}=20$ (GPU-B, correct
 center $(4,4)$), seven snapshots from $t=0.1$ to $t=2.0$.
 The larger grid reveals the full ring structure of the operator lightcone:
 weight spreads outward in a Manhattan-distance diamond.
 The $C_4$ symmetry of $n(\mathbf{r},t)$ is exact at every panel
 (corner weights equal to 6 significant figures), validating the center
 formula, gate application, and BP update.
 $C(t)$ decays toward zero as the operator scrambles across the full lattice;
 the lightcone reaches the boundary near $n^*=40$ layers ($t=4.0$), consistent
 with Table~\ref{tab:lightcone}.
}
\label{fig:lightcone_9x9}
\end{figure}

Figures~\ref{fig:lightcone_trap}--\ref{fig:lightcone_9x9} visualize
$n(\mathbf{r},t)$ as a spatial heatmap strip across seven Trotter snapshots,
covering three representative runs: the $4\times4$ trapped case at $\chi=80$,
the $4\times4$ clean baseline at $\chi=40$, and the $9\times9$ $C_4$-symmetric
showpiece at $\chi=20$.

Table~\ref{tab:ct_grids} compares $\Ct$ trajectories across grid sizes at
$\chi=50$ (GPU-B), demonstrating both the common early-time behavior and
the grid-dependent saturation.

\begin{table}[ht]
\centering
\caption{$\Ct$ at selected Trotter layers for $n\times n$ grids, $\chi=50$, GPU-B.
 All grids share the same $\Ct$ for $t \leq 0.9$ to 4 significant figures
 (early-time universality).
 Divergence begins when the operator front reaches the lattice boundary,
 consistent with the lightcone bound $n^*$ in Table~\ref{tab:lightcone}.
}
\label{tab:ct_grids}
\begin{tabular}{r rrrrrrrr}
\toprule
$t$ & $3\times3$ & $4\times4$ & $5\times5$ & $6\times6$ & $7\times7$ & $8\times8$ & $9\times9$ & $10\times10$ \\
\midrule
0.9 & 0.140 & 0.140 & 0.140 & 0.140 & 0.140 & 0.140 & 0.140 & 0.140 \\
1.4 & 0.113 & 0.085 & 0.096 & 0.096 & 0.096 & 0.096 & 0.096 & 0.096 \\
1.9 & 0.101 & 0.058 & 0.060 & 0.060 & 0.060 & 0.065 & 0.065 & 0.065 \\
2.4 & --- & --- & 0.050 & 0.050 & --- & --- & --- & --- \\
2.9 & --- & --- & 0.038 & 0.038 & 0.039 & 0.020 & 0.020 & 0.019 \\
3.9 & --- & --- & --- & --- & 0.002 & 0.000 & 0.000 & 0.000 \\
\midrule
\multicolumn{9}{l}{--- : beyond run length or beyond saturation.}\\
\bottomrule
\end{tabular}
\end{table}

Three regimes are visible in Table~\ref{tab:ct_grids}:
\begin{enumerate}
 \item \textbf{Early-time universality ($t \leq 0.9$)}: all grids agree, because
 the operator front has not yet reached any boundary regardless of grid size.
 \item \textbf{Boundary entry ($n^* \leq n \leq 2n^*$)}: $\Ct$ begins to depend
 on grid size as the front reflects from the boundaries.
 The $3\times3$ grid shows the strongest finite-size effects ($\Ct$ rises
 at $t \approx 0.9$ as the operator re-focuses after reflection).
 \item \textbf{Scrambling plateau ($n \gg n^*$)}: $\Ct \to 0$ as the operator
 distributes across all Pauli strings.
 The rate is set by the system size: larger grids support more scrambled
 configurations, so $\Ct$ decays more slowly.
\end{enumerate}

\begin{figure}[ht]
\centering
\includegraphics[width=\linewidth]{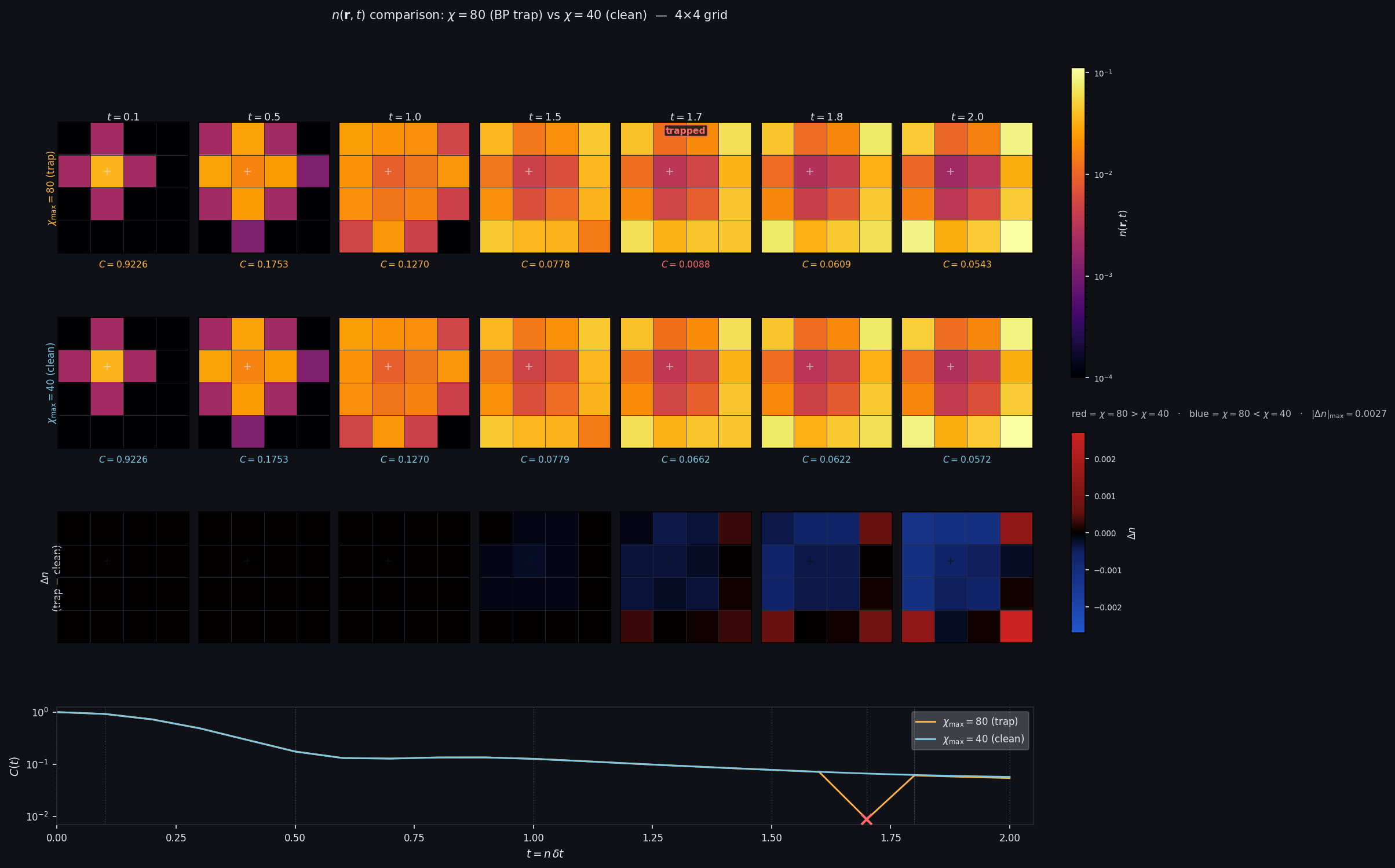}
\caption{%
 Spatial $\chi$-convergence on $4\times4$:
 $\Delta n(\mathbf{r},t) = n_{\chi=80} - n_{\chi=40}$ at six snapshots.
 Row~1 ($\chi=80$, BP trap): $n(\mathbf{r},t)$ on inferno log scale.
 Row~2 ($\chi=40$, clean): same scale.
 Row~3 ($\Delta n$): diverging colorscale, blue $\to$ black (zero) $\to$ red.
 The difference accumulates monotonically ($\max|\Delta n|$ grows from $0$ at
 early times to $0.0027$ at $t=2.0$), reflecting cumulative truncation
 differences.
 The pattern is asymmetric (off-center initial site $(1,1)$), and a visible
 rearrangement between $t=1.7$ (trapped, red $\times$) and $t=1.8$ (recovered)
 provides a spatial fingerprint of the BP metastable trap.
 Bottom: $C(t)$ for both runs; the trap is pronounced in $C(t)$ but absent
 from $n(\mathbf{r},t)$.
}
\label{fig:diff_4x4}
\end{figure}

\begin{figure}[ht]
\centering
\includegraphics[width=\linewidth]{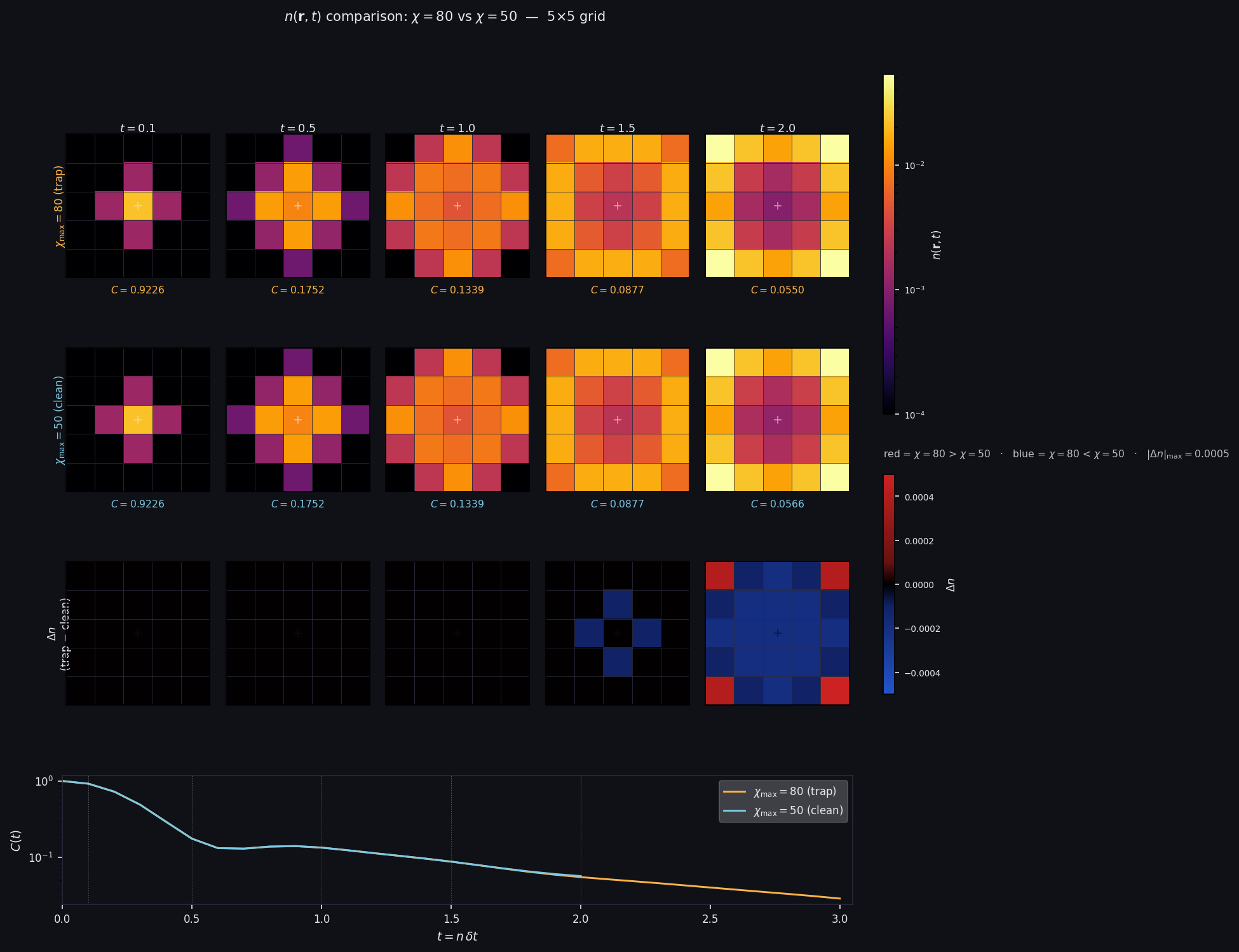}
\caption{%
 Spatial $\chi$-convergence on $5\times5$ (symmetric control):
 $\Delta n(\mathbf{r},t) = n_{\chi=80} - n_{\chi=50}$ at six snapshots.
 Same layout as Figure~\ref{fig:diff_4x4}.
 The symmetric center $(2,2)$ imposes exact $C_4$ symmetry on $\Delta n$
 at every panel (red interior ring, blue corners at late times), with no
 anomalous step at any layer.
 This confirms that the monotone $\max|\Delta n|$ growth on a symmetric grid
 is a pure bond-dimension effect, uncontaminated by the BP fixed-point
 multiplicity that distorts the $4\times4$ pattern.
 The pair of figures (Figs.~\ref{fig:diff_4x4}--\ref{fig:diff_5x5}) thus
 decomposes $\chi$-convergence of the spatial weights into a symmetric
 bulk contribution and an asymmetry-driven trap contribution.
}
\label{fig:diff_5x5}
\end{figure}

\paragraph{Spatial $\chi$-convergence: the difference field $\Delta n(\mathbf{r},t)$.}
The per-site weight $n(\mathbf{r},t)$ provides a richer convergence measure than
$C(t)$ alone: rather than a single scalar, it maps convergence onto the
two-dimensional lattice, revealing \emph{where} the bond-dimension difference
concentrates.
Figures~\ref{fig:diff_4x4} and~\ref{fig:diff_5x5} show the difference field
$\Delta n(\mathbf{r},t) = n_{\chi_1}(\mathbf{r},t) - n_{\chi_2}(\mathbf{r},t)$
for the $4\times4$ and $5\times5$ grids respectively.

On the \emph{symmetric} $5\times5$ grid (Figure~\ref{fig:diff_5x5},
$\chi_1=80$ versus $\chi_2=50$), $\Delta n$ is exactly $C_4$-symmetric at every
snapshot: a red interior ring (higher $\chi$ places more weight near the operator
origin) surrounded by blue corners (lower $\chi$ slightly overestimates corner
weight at late times).
The magnitude grows monotonically in time and no layer shows an anomalous step,
confirming that $\Delta n$ is a clean measure of bond-dimension convergence
on a symmetric grid, uncontaminated by fixed-point multiplicity.

On the \emph{asymmetric} $4\times4$ grid (Figure~\ref{fig:diff_4x4},
$\chi=80$ versus $\chi=40$, BP trap at $t=1.7$), the same monotone accumulation
is present but the pattern is \emph{not} $C_4$-symmetric: the off-center initial
site $(1,1)$ breaks the spatial symmetry of the difference field.
Between $t=1.7$ (trapped layer) and $t=1.8$ (recovered), a visible rearrangement
of the $\Delta n$ pattern appears --- a spatial fingerprint of the BP metastable
trap that is invisible in $n(\mathbf{r},t)$ alone but surfaces in $\Delta n$.
Taken together, the two figures establish that $n(\mathbf{r},t)$ converges in
$\chi$ faster than $C(t)$ (the maximum difference is ${<}0.3\%$ of peak weight),
and that the trap distortion, while detectable in $\Delta n$, does not qualitatively
alter the spatial spreading pattern.

\paragraph{Symmetric initial condition and BP stability.}
Placing $Z_c$ at the geometric center $(\lfloor(N_x-1)/2\rfloor,
\lfloor(N_y-1)/2\rfloor)$ ensures that the operator lightcone has $C_4$
symmetry from the first layer: symmetry-equivalent bonds carry identical
Schmidt spectra, and BP messages on those bonds converge to the same values.
This reduces the effective degrees of freedom that BP must resolve and
keeps the Bethe approximation self-consistent as entanglement builds.
Displacing $Z_c$ by even one site breaks this symmetry, creating an
asymmetric entanglement structure that stresses BP progressively as
$S_e/\log\chi$ grows.

The effect is directly measurable.
Table~\ref{tab:center_compare} compares BP sweep counts and Bethe norm
for the $9\times9$ grid at $\chi=20$ with the correct center $(4,4)$
and the off-center formula $N_x/2 - 1$ giving $(3,3)$.

\begin{table}[ht]
\centering
\caption{BP convergence and Bethe norm for $9\times9$, $\chi=20$: correct
 center $(4,4)$ versus off-center $(3,3)$.
 $S_e/\log\chi$ is identical for both runs (a property of $\chi=20$ on
 this grid) and included for reference.
 \textbf{Bold}: anomalous values.}
\label{tab:center_compare}
\small
\begin{tabular}{r r rr rr}
\toprule
 & & \multicolumn{2}{c}{Off-center $(3,3)$} & \multicolumn{2}{c}{Correct center $(4,4)$} \\
\cmidrule(lr){3-4}\cmidrule(lr){5-6}
Layer & $S_e/\log\chi$ & sweeps & norm & sweeps & norm \\
\midrule
15 & 0.505 & 11 & 1.000 & 11 & 1.000 \\
16 & 0.550 & 10 & 1.000 & 17 & 1.000 \\
17 & 0.597 & \textbf{36} & 1.000 & 18 & 1.000 \\
18 & 0.643 & 12 & 1.000 & 10 & 1.000 \\
19 & 0.686 & 24 & \textbf{0.001} & 12 & 1.000 \\
20 & 0.726 & 26 & \textbf{0.002} & 11 & 1.000 \\
21 & 0.761 & \textbf{100} & \textbf{5.095} & \textbf{100} & \textbf{1.000} \\
22 & 0.792 & 11 & \textbf{inf} & 10 & 1.000 \\
50 & 0.965 & 39 & \textbf{inf} & 36 & \textbf{1.005} \\
\bottomrule
\end{tabular}
\end{table}

Three observations follow from Table~\ref{tab:center_compare}.
First, $S_e/\log\chi$ crosses $0.5$ at layer 15 for both runs: the bond
saturation pressure is identical and is a property of $\chi=20$ on the
$9\times9$ grid, not of the center choice.
Second, the off-center run shows a first stress signal at layer 17
(36 sweeps versus 18 for the correct center), two layers before the
explosion — consistent with BP resolving a harder, asymmetric message
landscape under the same saturation pressure.
Third, at layer 21 both runs require 100 sweeps, but the outcomes are
qualitatively different: the off-center run explodes to norm $= 5.09$
and stays at \texttt{inf} for all remaining layers, while the correct
center stays at norm $= 1.000$ and drifts gently to $1.005$ by layer 50
--- a factor of $5000\times$ difference in Bethe norm deviation at a
single layer.

% ── Finite-size scaling ───────────────────────────────────────────────────────
\subsection{Finite-Size Scaling and Bethe Approximation Stability}
\label{sec:finite_size}

As the grid grows, two effects compound: more closed loops in the lattice
make the Bethe approximation less accurate (increasing the effective
$\chi^*$ needed for a given gate error), and the larger variational space
requires more BP sweeps to converge.
We characterize both effects using the Bethe norm
$\|\psi\|^2_{\mathrm{Bethe}} = \exp(\sum_v \log f_v - \sum_e \log Z_e)$,
which equals 1 under the exact Bethe approximation and deviates when
the approximation breaks down.

\begin{table}[ht]
\centering
\caption{Finite-size scaling of Bethe approximation stability.
 $\chi=50$ on GPU-B, except $n\times n \geq 8$ where $\chi=20$.
 ``Norm stable'' means Bethe norm $< 2.0$ throughout the run.
 $n^*$: lightcone saturation layer (Table~\ref{tab:lightcone}).
 gate\_err at $n^*$ is the accumulated truncation error at saturation.}
\label{tab:finite_size}
\begin{tabular}{cccccc}
\toprule
Grid & $\chi$ & Steps & Norm stable & gate\_err at $n^*$ & $\Ct$ at $n^*$ \\
\midrule
$3\times3$ & 50 & 20 & \checkmark & $7\times10^{-5}$ & 0.111 \\
$4\times4$ & 50 & 20 & \checkmark & $5\times10^{-4}$ & 0.056 \\
$5\times5$ & 50 & 30 & \checkmark & $2\times10^{-3}$ & 0.034 \\
$6\times6$ & 50 & 30 & \checkmark & $7\times10^{-3}$ & 0.033 \\
$7\times7$ & 60 & 40 & \checkmark & $1\times10^{-2}$ & 0.002 \\
$8\times8$ & 20 & 40 & \checkmark$^\dagger$ & $6\times10^{-2}$ & 0.000 \\
$9\times9$ & 20 & 50 & \checkmark & $>10^{-1}$ & 0.000 \\
$10\times10$ & 20 & 50 & \checkmark$^\dagger$ & $>10^{-1}$ & 0.000 \\
\midrule
\multicolumn{6}{l}{$^\dagger$ Norm exceeds 2.0 transiently but recovers; $\Ct$ is unaffected.}\\
\bottomrule
\end{tabular}
\end{table}

The gate error at saturation grows steadily with grid size at fixed $\chi$,
reflecting the increasing difficulty of maintaining the Bethe approximation
on more densely connected lattices.
For $5\times5$ and $6\times6$ at $\chi=50$, gate\_err is marginal
($\sim 2$--$7\times10^{-3}$) but $\Ct$ plateaus smoothly, indicating that
the truncation error has not corrupted the correlation function.
By $7\times7$, $\chi=50$ is insufficient (gate\_err $> 10^{-2}$) and $\chi=60$
is needed; at $8\times8$ and beyond, $\chi=20$ is the feasibility ceiling on
GPU-B for multi-step runs, and gate\_err signals that $\Ct$ is qualitative rather
than quantitative.

A striking result is the $10\times10$ run at $\chi=20$:
despite gate\_err $\approx 0.1$ and large Bethe norm fluctuations,
$\Ct$ follows the same qualitative trajectory as smaller grids
(rapid decay to $\approx 0$ by $t = 4.0$), confirming that the scrambling
physics is captured even when the variational approximation is far from exact.
This demonstrates feasibility of the $10\times10$ geometry on GPU-B at $\chi=20$
and motivates the primary research target: $10\times10$ at $\chi=100$
(requiring $\approx109$\,GB, see §\ref{sec:memory_model}).

% ── Cross-architecture portability (physics) ──────────────────────────────────
\subsection{Cross-Configuration Portability: Physics Comparison}
\label{sec:portability_physics}

The GPU-A runs provide the first systematic comparison of
$\Ct$ across the two 32\,GB configurations for $\chi \in \{40, 60, 110\}$
and grid sizes $5\times5$ through $7\times7$.

\paragraph{$4\times4$: portable operating points.}
Table~\ref{tab:mi60_chi} compares $\Ct$ at $t=2.0$ for both architectures.

\begin{table}[ht]
\centering
\caption{$\Ct$ at $t = 2.0$ on $4\times4$: GPU-B versus GPU-A, all $\chi$.
 \textbf{Bold}: metastable BP trap.
 Daggers mark the two portable operating points where both architectures agree
 and are trap-free.}
\label{tab:mi60_chi}
\begin{tabular}{cllr}
\toprule
$\chi$ & GPU-B & GPU-A & $|\Delta \Ct|$ \\
\midrule
20 & 0.0646 (trap L16) & 0.0631 (trap L17) & 0.0015 \\
40 & 0.0572$^\dagger$ & 0.0577$^\dagger$ & 0.0005 \\
50 & 0.0558 & \textbf{0.0073} & --- \\
60 & 0.0551$^\dagger$ & 0.0546$^\dagger$ & 0.0005 \\
70 & \textbf{0.0072} & 0.054 (est.) & --- \\
80 & 0.0543 & \textbf{0.0073} & --- \\
100 & \textbf{0.0071} & 0.054 (est.) & --- \\
110 & 0.0537 & BLAS fail & --- \\
\midrule
\multicolumn{4}{l}{$^\dagger$ Portable operating points: trap-free on both architectures,
 $|\Delta\Ct| < 10^{-3}$.}\\
\bottomrule
\end{tabular}
\end{table}

The results exhibit perfect complementarity in the trap pattern:
GPU-A traps at $\chi \in \{50, 80\}$ while GPU-B is clean;
GPU-B traps at $\chi \in \{70, 100\}$ while GPU-A is clean.
Neither architecture is strictly more reliable than the other;
the trap patterns are fingerprints of the BP fixed-point landscape
selected by the matrix multiply-accumulate reduction tree (GPU-B) versus scalar vector-ALU (GPU-A).

At $\chi=40$ and $\chi=60$, both architectures are trap-free and agree
to within $5\times10^{-4}$ --- better than $0.1\%$.
These are the \emph{portable} bond dimensions for this system.

\paragraph{GPU-A ceiling: BLAS coverage and BP.}
At $\chi=110$, the GPU-A run fails for two compounding reasons.
First, BP enters a deep trap at $t=0.7$ ($\Ct = 8.7\times10^{-5}$,
sweeps $= 100$), indicating that the BP fixed-point landscape at this
bond dimension is already problematic.
Second, at the GEMM dimensions arising at $\chi=110$
($m = \chi^3 \approx 1.3\times10^6$, $n = D\chi = 440$),
GPU-A lacks pre-built BLAS kernels for these matrix dimensions and the fallback
runtime compilation fails.
The GPU-A ceiling is $\chi \lesssim 100$ for correctness and
$\chi \lesssim 80$ for reliable BP convergence.

\paragraph{Grid scaling on GPU-A.}
For $5\times5$ through $7\times7$ at $\chi=50$, the GPU-A runs with corrected
centers produce $\Ct$ trajectories that are physically consistent with the
GPU-B odd\_fix results.
Table~\ref{tab:mi60_grids} shows $\Ct$ at saturation.

\begin{table}[ht]
\centering
\caption{$\Ct$ at $n^*$ (lightcone saturation layer) for odd grids,
 $\chi=50$, GPU-B versus GPU-A (both with corrected geometric center).
 Agreement confirms that grid-size physics is portable at $\chi=50$.}
\label{tab:mi60_grids}
\begin{tabular}{ccrrrr}
\toprule
Grid & $n^*$ & Steps & GPU-B $\Ct(n^*)$ & GPU-A $\Ct(n^*)$ & $|\Delta\Ct|$ \\
\midrule
$5\times5$ & 20 & 30 & 0.057 & 0.057 & $<0.001$ \\
$6\times6$ & 30 & 30 & 0.038 & 0.033 & 0.005 \\
$7\times7$ & 30 & 30 & 0.039 & 0.033 & 0.006 \\
\bottomrule
\end{tabular}
\end{table}

The $5\times5$ result is portable to within the noise floor.
The $6\times6$ and $7\times7$ discrepancies ($\sim 0.005$) grow with
grid size, consistent with the larger number of BP update paths and the
growing sensitivity of the fixed-point selection to floating-point order.
At these grid sizes, the dominant source of uncertainty in $\Ct$ is the
BP fixed-point selection, not the truncation error.

% ── Grid scaling (timing) — keep existing text ────────────────────────────────
\FloatBarrier % flush lightcone/diff figures before the timing table
\subsection{Grid Scaling (Timing)}
\label{sec:grid_timing}

\begin{figure}[ht]
\centering
\small
\begin{tabular}{ccrrrrrr}
\toprule
Grid & Bonds & \multicolumn{2}{c}{GPU-B (ms)} & \multicolumn{2}{c}{GPU-A (ms)} & GPU-A/GPU-B \\
\cmidrule(lr){3-4}\cmidrule(lr){5-6}
 & & Adaptive & Householder & Adaptive & Householder & (HH) \\
\midrule
$3\times3$ & 12 & 2{,}167 & 2{,}678 & 10{,}656 & 11{,}935 & 4.5$\times$ \\
$4\times4$ & 24 & 7{,}748 & 8{,}274 & 31{,}453 & 37{,}323 & 4.5$\times$ \\
$5\times5$ & 40 & 12{,}802 & 16{,}316 & 91{,}044 & 83{,}090 & 5.1$\times$ \\
$6\times6$ & 60 & 21{,}270 & 23{,}332 & 135{,}010 & 118{,}046 & 5.1$\times$ \\
$7\times7$ & 84 & 25{,}219 & 31{,}208 & 138{,}317 & 156{,}576 & 5.0$\times$ \\
$8\times8$ & 112 & 32{,}880 & 36{,}654 & 207{,}509 & 198{,}907 & 5.4$\times$ \\
\bottomrule
\end{tabular}
\bigskip

\includegraphics[width=0.75\linewidth]{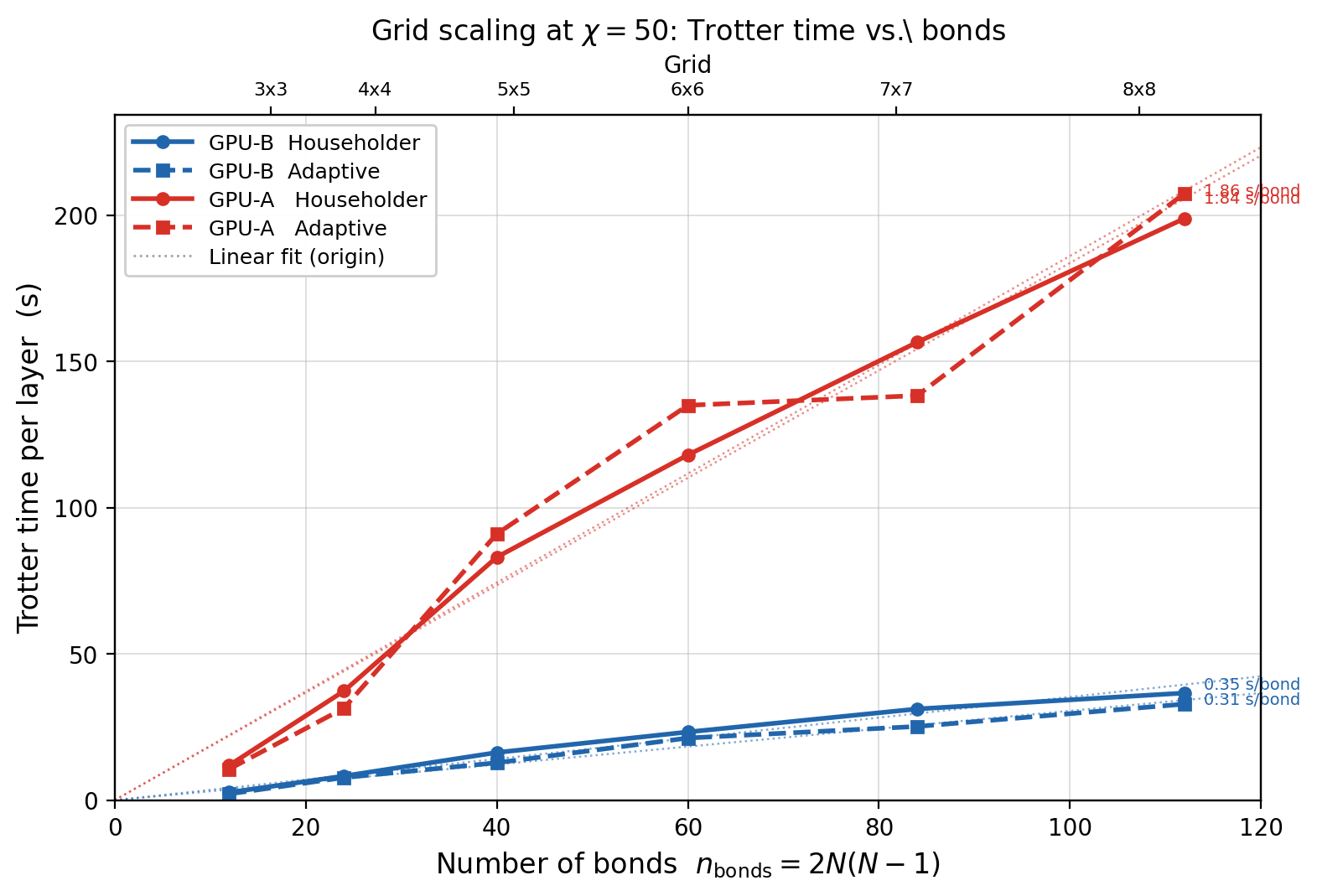}
\caption{\small Grid scaling at $\chi=50$: Trotter time per layer (average over all layers)
 for square grids on GPU-B and GPU-A, both QR modes.
 \emph{Top}: raw times in ms; bonds $= 2N(N-1)$ for an $N\times N$ grid;
 GPU-A/GPU-B ratio computed from Householder columns.
 \emph{Bottom}: same data plotted versus bond count.
 Solid lines ($\circ$): Householder QR; dashed lines ($\square$): Adaptive QR;
 blue: GPU-B; red: GPU-A; dotted: linear fits through the origin
 (slopes annotated in s/bond).
 Both GPU-B modes scale linearly with bond count, confirming that gate cost
 depends only on $\chi$ and site degree, not grid size.
 Adaptive is 10--20\% faster on GPU-B because Cholesky avoids the full
 Householder factorization when the Gram matrix is well-conditioned.
 GPU-A is approximately $5\times$ slower than GPU-B across all grid sizes,
 consistent with the matrix multiply-accumulate versus vector-FP throughput ratio at $\chi=50$.}
\label{tab:grid_scaling}\label{fig:grid_scaling}
\end{figure}

Gate cost is independent of grid size (depends only on $\chi$ and site degree).
Trotter time per layer scales linearly with $n_{\text{bonds}}$, as confirmed in
Table~\ref{tab:grid_scaling}: the ratio to $3\times3$ closely tracks
$n_{\text{bonds}}/12$ for both architectures and both QR modes,
with at most 15\% deviation at the largest grids attributable to cache
and memory-bandwidth pressure.

% ── Memory model — keep existing text ────────────────────────────────────────
\FloatBarrier
\subsection{Memory Model Validation}
\label{sec:memory_validation}

The analytic model of Section~\ref{sec:memory_model} predicts peak VRAM from
$\chimax$ and the grid dimensions alone, before the simulation runs.
Table~\ref{tab:vram} compares predictions against \texttt{rocm-smi} measurements
taken at the start of each run on a $4\times4$ grid.
The model is accurate to within 2\% for $\chimax \leq 100$, confirming that
the dominant terms in equations~\eqref{eq:mem_psi}--\eqref{eq:mem_ws} correctly
account for the pre-allocated psi tensors, messages, and workspace.
The chi=110 run on GPU-B was measured at 94\% VRAM utilization
($\approx 30.1$\,GB of 32\,GB), confirming the analytic prediction
of $\approx 30$\,GB (Table~\ref{tab:vram}).
The remaining 6\% overhead is GPU driver and context allocation.
This establishes $\chimax = 110$ as the confirmed ceiling on a 32\,GB GPU-B
for a $4\times4$ grid; the analytic model predicts $\chimax = 108$ as the
limit, with the 2-unit gap explained by driver overhead not in the model.
This accuracy is sufficient for deployment planning: given a target grid and
$\chimax$, the model determines whether a GPU can run the simulation before any
code is compiled or executed.
Table~\ref{tab:vram_grid} extends the predictions to larger grids and bond
dimensions, identifying the GPU generation required for each configuration.

\begin{figure}[ht]
\centering
\captionof{table}{Memory model validation: predicted versus measured VRAM on $4\times4$ grid
 (GPU-B and GPU-A, 32\,GB HBM). Prediction uses
 equations~\eqref{eq:mem_psi}--\eqref{eq:mem_ws}.}
\label{tab:vram}
\smallskip
\begin{tabular}{rrrr}
\toprule
$\chimax$ & Predicted (GB) & Measured VRAM & Error \\
\midrule
 50 & 0.82 & $\approx 3\%$ (1.0\,GB) & $< 1\%$ \\
 80 & 5.31 & 16\% (5.1\,GB) & $< 1\%$ \\
100 & 12.94 & 38\% (12.2\,GB) & $< 2\%$ \\
110 & 30.1 & 94\% (30.1\,GB) & $< 2\%$ \\
\midrule
\multicolumn{4}{l}{GPU-B 32\,GB; chi=110 measured at run completion.} \\
\bottomrule
\end{tabular}
\bigskip
\captionof{table}{Predicted peak VRAM (GB) for $n\times n$ grids at selected bond dimensions,
 from the analytic model (\S\ref{sec:memory_model}).
 $\bullet$~fits on GPU-B/GPU-A ($\leq 32$\,GB);
 $\circ$~requires more than 32\,GB.}
\label{tab:vram_grid}
\smallskip
\begin{tabular}{r rrrr}
\toprule
Grid & $\chi=50$ & $\chi=70$ & $\chi=90$ & $\chi=100$ \\
\midrule
$4\times4$ & 0.8\,$\bullet$ & 3.1\,$\bullet$ & 8.5\,$\bullet$ & 12.9\,$\bullet$ \\
$5\times5$ & 1.3\,$\bullet$ & 5.1\,$\bullet$ & 13.8\,$\bullet$ & 21.0\,$\bullet$ \\
$6\times6$ & 2.0\,$\bullet$ & 7.8\,$\bullet$ & 21.2\,$\bullet$ & 32.3\,$\circ$ \\
$7\times7$ & 2.9\,$\bullet$ & 11.3\,$\bullet$ & 30.7\,$\bullet$ & 46.7\,$\circ$ \\
$8\times8$ & 4.1\,$\bullet$ & 15.5\,$\bullet$ & 42.3\,$\circ$ & 64.4\,$\circ$ \\
$9\times9$ & 5.4\,$\bullet$ & 20.5\,$\bullet$ & 56.0\,$\circ$ & 85.3\,$\circ$ \\
$10\times10$ & 6.9\,$\bullet$ & 26.3\,$\bullet$ & 71.8\,$\circ$ & 109.3\,$\circ$$^\dagger$ \\
\midrule
\multicolumn{5}{l}{$^\dagger$ Primary research target.} \\
\bottomrule
\end{tabular}
\end{figure}

% ══════════════════════════════════════════════════════════════════════════════

% ══════════════════════════════════════════════════════════════════════════════
\section{Performance}
\label{sec:performance}
% ══════════════════════════════════════════════════════════════════════════════

We characterize performance along two axes: how Trotter and BP time each scale
with bond dimension $\chi$, and how CppSim compares to the Julia reference.

At $\chi=50$, $4\times4$ grid, GPU-B, CppSim achieves a Trotter time of
$\approx 8{,}300$\,ms/layer (Householder) versus $\approx 20{,}000$\,ms/layer
for the Julia/GPU reference --- a $\mathbf{2.4\times}$ speedup --- coming
primarily from the GPU permutation kernel (\S\ref{sec:permute}) and
strided-batched GEMM for environment absorption.
BP cost is higher in CppSim ($\approx 1{,}100$\,ms) than in Julia
($\approx 8$\,ms); CppSim runs up to 100 sweeps to full convergence
whereas the Julia reference caps at 10 sweeps, and the two implementations
may differ in message representation and convergence criterion.

\begin{figure}[p]
\centering
\includegraphics[width=\linewidth]{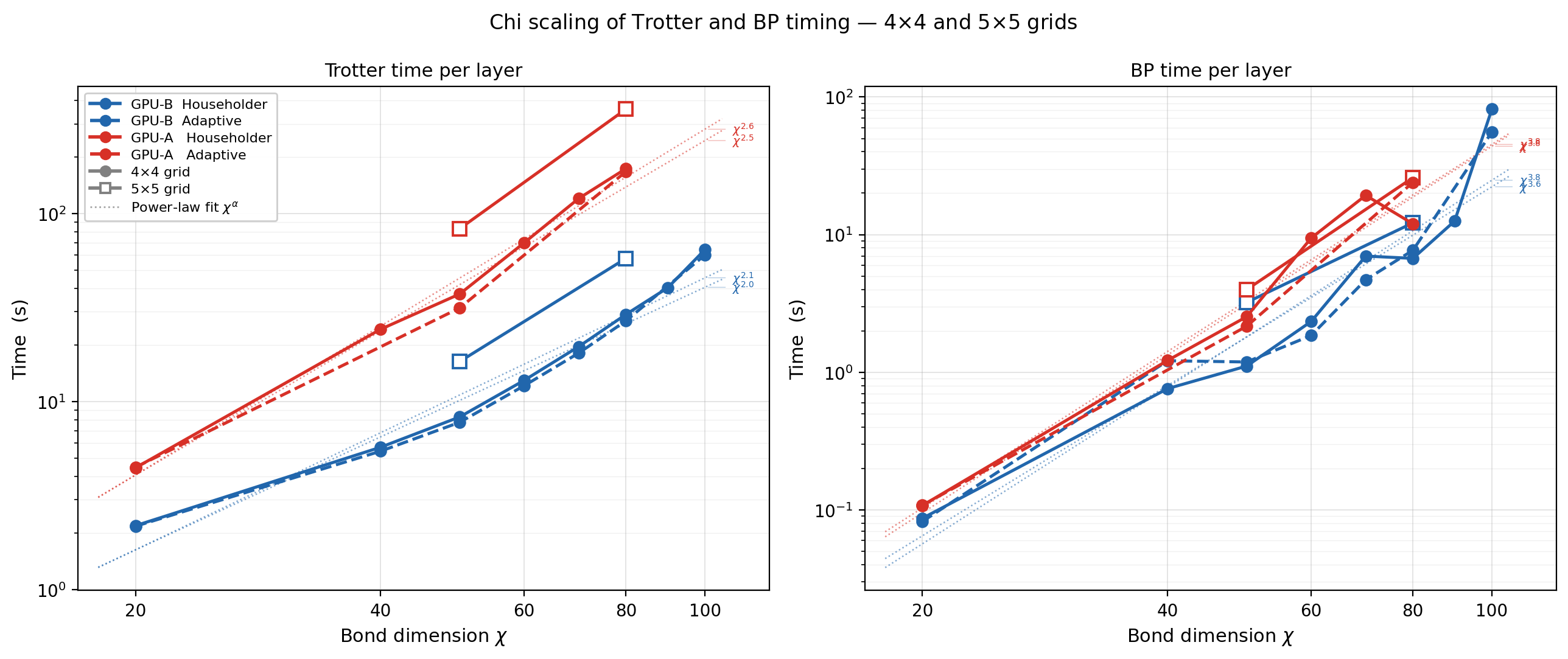}
\vspace{0.5em}
\includegraphics[width=0.88\linewidth]{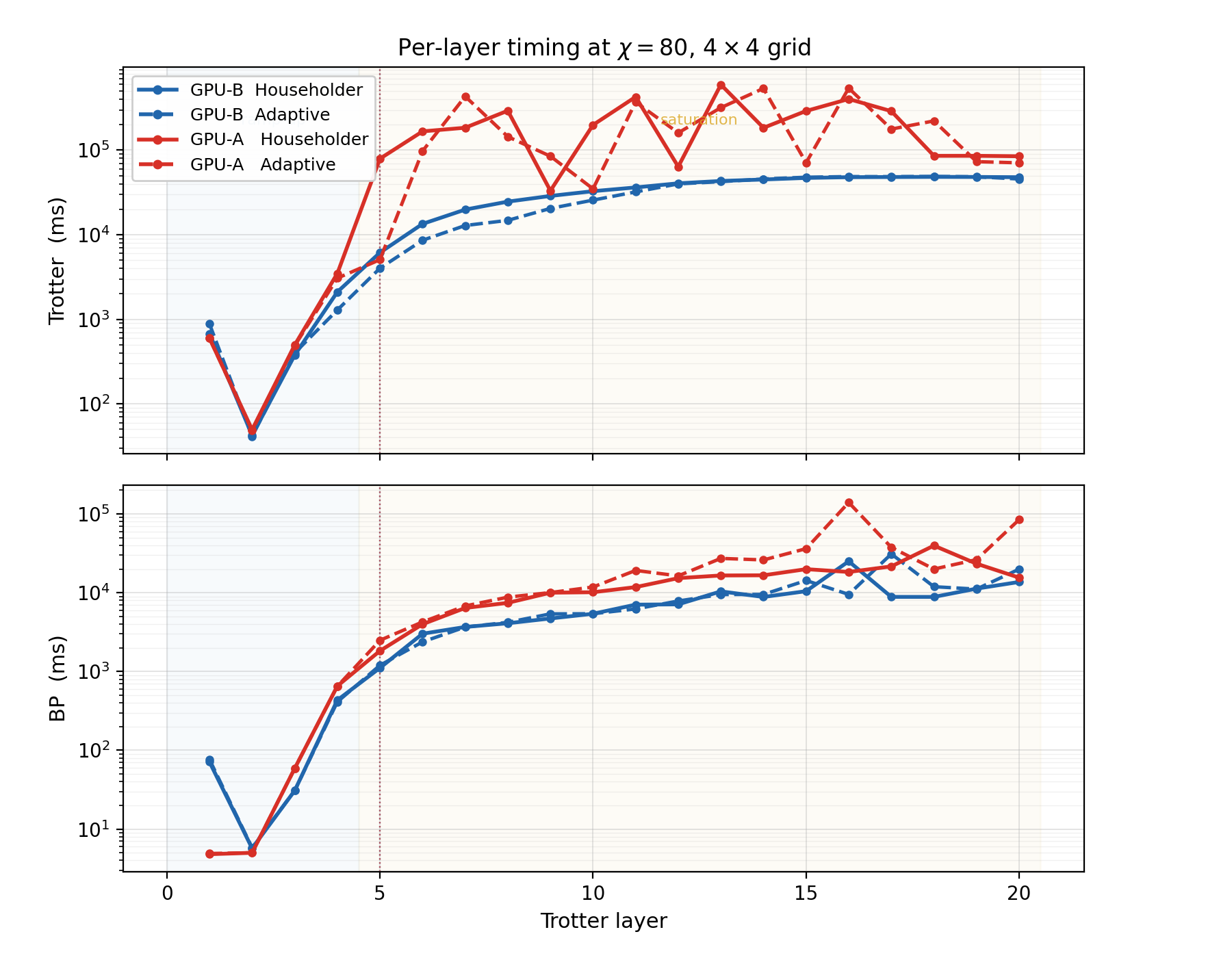}
\caption{%
 Performance at $\chi=80$, $4\times4$ grid, in full detail.
 \emph{Top}: Trotter and BP time averaged per layer across the full $\chi$ range
 (log-log), showing how cost scales with bond dimension for GPU-B and GPU-A
 in both QR modes.
 \emph{Bottom}: the same two quantities layer by layer at $\chi=80$,
 revealing the ramp-up phase (blue shading), saturation plateau, and
 BP spikes at the trapped layers identified in Table~\ref{tab:trap_inventory}.}
\label{fig:timing_chi}\label{fig:breakdown_chi80}
\end{figure}

\clearpage
% ══════════════════════════════════════════════════════════════════════════════
\section{Memory-Bandwidth Regime and Numerical Precision}
\label{sec:portability}
% ══════════════════════════════════════════════════════════════════════════════

The dominant operations --- SVD, QR, and environment absorption GEMMs ---
are memory-bandwidth-bound at the matrix shapes arising from $\chi \leq 100$:
the arithmetic intensity falls well below the ridge point of any modern GPU,
so FLOP count alone does not predict performance.
To quantify this, we ran on two configurations with identical VRAM (32\,GB)
but different characteristics: GPU-A (1.0\,TB/s, 29.5\,TFLOPS) and GPU-B
(1.2\,TB/s, 184.6\,TFLOPS).
The observed wall-clock ratio ($\approx5\times$, Figure~\ref{fig:timing_chi})
lies well below the compute ratio ($6.3\times$), confirming that compute
throughput is not the limiting resource.
The practical consequence for algorithm selection is deferred to the
high-bandwidth regime (Appendix~\ref{app:mi355x}), where the difference
becomes visible.

\paragraph{Float32 accumulation error in Cholesky-QR.}
A subtler algorithmic finding concerns numerical precision.
At $\chimax=100$, results from the two GPU configurations drift by
$1$--$3\times10^{-3}$ at late times ($t \geq 12$), growing monotonically
across layers. This drift is absent at $\chimax \leq 80$ and is not caused
by BP sign anomalies (which recover within one layer at all $\chi$).

The root cause is the gram matrix GEMM in the Cholesky-QR path
(\S\ref{sec:qr}): $\mathbf{C} = \mathbf{cv}^\top \mathbf{cv}$ is a
float32 reduction over $k = \chimax^3$ terms ($k = 10^6$ at $\chi=100$).
Since floating-point addition is non-associative, different SGEMM
implementations --- scalar vector-ALU versus matrix-multiply units with
different reduction trees --- produce different rounding errors.
The effect grows as $k = \chi^3$, consistent with the observed
$\chi$-dependence of the drift.
The Julia reference uses the same Cholesky-QR algorithm but in float64,
making the gram matrix computation essentially exact regardless of
summation order.

% ══════════════════════════════════════════════════════════════════════════════
\section{Conclusion}
\label{sec:conclusion}
% ══════════════════════════════════════════════════════════════════════════════

We have presented \textsc{CppSim}, a C++/GPU implementation of Heisenberg-picture
tensor network dynamics for the 2D Ising model on GPU hardware.
\textsc{CppSim} runs bond dimensions up to $\chi=110$ on a single 32\,GB GPU,
a scale not previously reported for this model.
The validated memory model projects the primary research target ---
$10\times10$ at $\chi=100$, requiring $\approx109$\,GB --- within reach of
high-memory accelerators, opening a new scale of 2D quantum dynamics simulation.

\paragraph{Outlook.}
Beyond the immediate results, this work demonstrates that tensor network simulation
at large bond dimension is a memory-bandwidth problem, not a compute problem:
efficiency gains come from HBM bandwidth rather than FLOP count; real-number
representation in the Pauli basis halves VRAM over complex storage; and on-device
computation eliminates PCIe as the remaining bottleneck.
Float32 precision itself becomes architecture-dependent at large $\chi$, pointing
toward mixed-precision accumulation as a necessary step for reproducible results
across GPU generations.
The remaining SVD synchronisation point (\S\ref{sec:performance}) currently
prevents stream-level parallelism; replacing it with an on-device singular-value
reduction would enable four concurrent gate streams on a $4\times4$ grid ---
one per independent colour of the Trotter checkerboard --- a multiplier that
grows with grid size and is best realised on high-VRAM hardware.
These are not \textsc{CppSim}-specific conclusions --- they are design principles
at the intersection where physicists choose what to simulate and computer scientists
choose how.

\paragraph{Availability.}
Source code, build instructions, and experiment scripts are available from
the corresponding author upon request.

% ── Bibliography ──────────────────────────────────────────────────────────────
\bibliographystyle{abbrvnat}

\begin{thebibliography}{99}

\bibitem{orus2014practical}
R.~Or\'us.
\newblock A practical introduction to tensor networks: Matrix product states
 and projected entangled pair states.
\newblock \emph{Annals of Physics}, 349:117--158, 2014.

\bibitem{vidal2003efficient}
G.~Vidal.
\newblock Efficient classical simulation of slightly entangled quantum
 computations.
\newblock \emph{Physical Review Letters}, 91(14):147902, 2003.

\bibitem{vidal2004efficient}
G.~Vidal.
\newblock Efficient simulation of one-dimensional quantum many-body systems.
\newblock \emph{Physical Review Letters}, 93(4):040502, 2004.
\newblock \doi{10.1103/PhysRevLett.93.040502}.

\bibitem{rocblas}
{GPU runtime}.
\newblock GPU BLAS library.
\newblock \url{https://github.com/GPU runtimeSoftwarePlatform/BLAS}, 2024.

\bibitem{rocsolver}
{GPU runtime}.
\newblock GPU LAPACK library.
\newblock \url{https://github.com/GPU runtimeSoftwarePlatform/LAPACK}, 2024.

\bibitem{sachdev2011quantum}
S.~Sachdev.
\newblock \emph{Quantum Phase Transitions}.
\newblock Cambridge University Press, Cambridge, 2nd edition, 2011.
\newblock \doi{10.1017/CBO9780511973765}.

\bibitem{suzuki1976generalized}
M.~Suzuki.
\newblock Generalized {Trotter}'s formula and systematic approximants of
 exponential operators.
\newblock \emph{Communications in Mathematical Physics}, 51(2):183--190, 1976.

\bibitem{golub2013matrix}
G.~H.~Golub and C.~F.~Van~Loan.
\newblock \emph{Matrix Computations}.
\newblock Johns Hopkins University Press, 4th edition, 2013.
\newblock ISBN 9781421407944.

\bibitem{rudolph2025simulating}
M.~S.~Rudolph and J.~Tindall.
\newblock Simulating and sampling from quantum circuits with {2D} tensor
 networks.
\newblock \emph{arXiv preprint arXiv:2507.11424}, 2025.
\newblock \url{https://arxiv.org/abs/2507.11424}.

\bibitem{tindall2023gauging}
J.~Tindall and M.~Fishman.
\newblock Gauging tensor networks with belief propagation.
\newblock \emph{SciPost Physics}, 15:222, 2023.
\newblock \doi{10.21468/SciPostPhys.15.6.222}.

\bibitem{yedidia2005constructing}
J.~S.~Yedidia, W.~T.~Freeman, and Y.~Weiss.
\newblock Constructing free-energy approximations and generalized belief
 propagation algorithms.
\newblock \emph{IEEE Transactions on Information Theory}, 51(7):2282--2312,
 2005.
\newblock \doi{10.1109/TIT.2005.850085}.

\bibitem{liebrobinson}
E.~H.~Lieb and D.~W.~Robinson.
\newblock The finite group velocity of quantum spin systems.
\newblock \emph{Communications in Mathematical Physics}, \textbf{28}:251--257,
 1972.

\end{thebibliography}

% ── Appendices ────────────────────────────────────────────────────────────────
\appendix

% ── GPU-C appendix ───────────────────────────────────────────────────────────
% ══════════════════════════════════════════════════════════════════════════════
\section{High-Bandwidth Memory Regime: Adaptive QR and Large-Scale Simulations}
\label{app:mi355x}
% ══════════════════════════════════════════════════════════════════════════════

Section~\ref{sec:portability} established that \textsc{CppSim} operates in the
memory-bandwidth-bound regime at $\chi \leq 100$ on $4\times4$ grids, where
algorithm choice (Householder vs.\ Cholesky-QR) is invisible in wall-clock time.
This appendix reports what happens when bandwidth is no longer the bottleneck:
a GPU configuration with $5.5\times$ higher memory bandwidth and 192+\,GB VRAM
(GPU-C) shifts the workload into the
compute-bound regime at larger grids, making the Cholesky-QR advantage measurable.
The large VRAM also enables the primary research target --- $10\times10$ at
$\chimax=100$, requiring $\approx109$\,GB.

\subsection{Grid-scaling Trotter timing at $\chi = 50$}

Table~\ref{tab:grid_scaling_mi355x} reports saturated-layer Trotter timing
as ratios relative to each configuration's own $3\times3$ Householder baseline,
so only within-configuration scaling and QR-mode differences are claimed.

\begin{table}[H]
\centering
\caption{Trotter time per layer at $\chi=50$, saturated layers only, normalized
 to the $3\times3$ Householder time within each configuration.
 \texttt{no\_sat}: not all bonds reached $\chi=50$;
 \texttt{---}: run not collected.}
\label{tab:grid_scaling_mi355x}
\smallskip
\small
\begin{tabular}{r r cc cc}
\toprule
& & \multicolumn{2}{c}{GPU-B (ref = $3\times3$ HH)} & \multicolumn{2}{c}{GPU-C (ref = $3\times3$ HH)} \\
\cmidrule(lr){3-4}\cmidrule(lr){5-6}
Grid & Bonds & HH & Adap & HH & Adap \\
\midrule
$3\times3$ & 12 & 1.00 & 1.04 & 1.00 & 0.78 \\
$4\times4$ & 24 & 3.72 & 3.55 & 3.67 & 2.76 \\
$5\times5$ & 40 & 7.57 & 6.86 & 7.62 & 5.55 \\
$6\times6$ & 60 & 12.45 & 11.33 & 12.98 & 8.39 \\
$7\times7$ & 84 & 18.18 & no\_sat & 19.57 & 13.20 \\
$8\times8$ & 112 & no\_sat & no\_sat & 27.71 & --- \\
\bottomrule
\end{tabular}
\end{table}

Both configurations scale identically under Householder QR across all grid
sizes, confirming consistent normalization and the same physical workload.
The \texttt{no\_sat} entries reflect a VRAM ceiling, not a compute or
algorithm limitation.

Cholesky-QR (Adap) tracks Householder within 10\% on GPU-B at all grid sizes,
but progressively pulls ahead on GPU-C --- from 22\% at $3\times3$ to 35\% at
$6\times6$. Sufficient memory bandwidth exposes Cholesky's lower arithmetic
constant; below that threshold, both QR modes are memory-traffic-equivalent.

\begin{figure}[H]
\centering
\includegraphics[width=\linewidth]{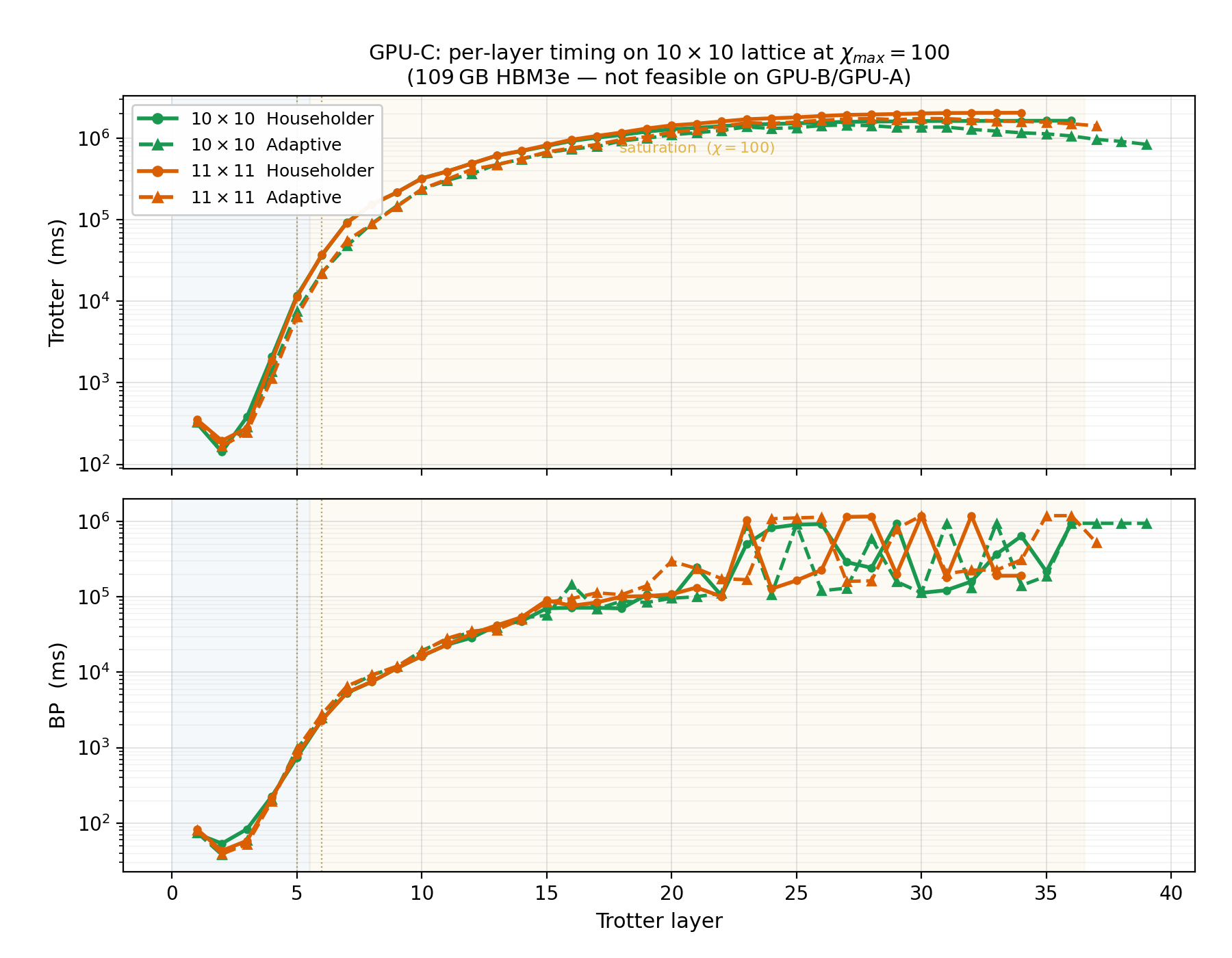}
\caption{Per-layer Trotter and BP time on the $10\times10$ lattice at
 $\chimax=100$, GPU-C only (high-bandwidth memory, 192+\,GB).
 This run requires $\approx 109$\,GB and is enabled by the 192+\,GB configuration.
 Solid line: Householder QR; dashed: Adaptive QR.
 Trotter time (top) grows smoothly as bond dimension saturates then climbs
 with increasing BP sweep count.
 BP time (bottom) is variable because convergence at $\chi=100$ requires
 up to 100 sweeps (the configured maximum); the spread reflects per-layer
 variation in BP fixed-point difficulty.
 Runs completed 40 of 100 requested layers within the 16\,h wall-time allocation.}
\label{fig:mi355x_10x10_chi100}
\end{figure}

\subsection{BP trap behaviour on GPU-C}

Table~\ref{tab:traps_mi355x} records Type-1 trap occurrences detected by
\texttt{detect\_traps.py} on the GPU-C grid-scaling logs at $\chi=50$.

\begin{table}[H]
\centering
\caption{Type-1 BP traps on GPU-C at $\chi=50$, grid-scaling suite.
 ``none'' = zero Type-1 events in the full run.}
\label{tab:traps_mi355x}
\smallskip
\small
\begin{tabular}{r cc}
\toprule
Grid & HH trap layers & Adap trap layers \\
\midrule
$3\times3$ & 15, 16 & 30 \\
$4\times4$ & 19 & none \\
$5\times5$ & none & none \\
$6\times6$ & none & none \\
$7\times7$ & none & none \\
$8\times8$ & none & --- \\
\bottomrule
\end{tabular}
\end{table}

The pattern is consistent with Table~\ref{tab:trap_inventory}: grids of
$5\times5$ and larger with odd dimensions are trap-free, confirming C4
lattice symmetry as the dominant determinant.
Small grids ($3\times3$, $4\times4$) trap at early layers where all bonds
saturate simultaneously; all Type-1 traps recover within one layer.
Trap layers differ by a few positions across configurations, consistent with
floating-point reduction order selecting a different but equivalent BP path.

\subsection{Spatial operator spreading at $10\times10$}

All $\chi=100$ runs used a 16\,h wall-time allocation, completing ${\approx}38$
Trotter steps (${\approx}70$ for $11\times11$).
These are the first tensor-network simulations of the Heisenberg-picture
Ising model on $10\times10$ and $11\times11$ lattices at $\chi=100$.

At sufficiently long times $C(t)$ decays to zero and the spatial distribution
$n(\mathbf{r},t)$ becomes the primary observable.
We characterize spreading using the per-site non-identity Pauli weight
$n(\mathbf{r},t)$ introduced in \S\ref{sec:lightcone}.
From it we extract two radii --- the mean Manhattan radius $R_{L1}(t)$ and RMS
Euclidean radius $R_{L2}(t)$ from the center site $\mathbf{r}_c$ --- and the
spatial entropy fraction
\begin{equation}
 S_{\rm frac}(t) = \frac{-\sum_{\mathbf{r}} p(\mathbf{r},t)\ln p(\mathbf{r},t)}
 {\ln N^2},
 \quad p(\mathbf{r},t) = \frac{n(\mathbf{r},t)}{\sum_{\mathbf{r}} n(\mathbf{r},t)},
 \label{eq:sfrac}
\end{equation}
the spatial analogue of the bond saturation fraction $S_e/\log\chi$
(\S\ref{sec:experiments}), normalized to $[0,1]$: zero for a perfectly localized
operator, one for uniform weight across all $N^2$ sites.

\begin{figure}[H]
\centering
\includegraphics[width=\linewidth]{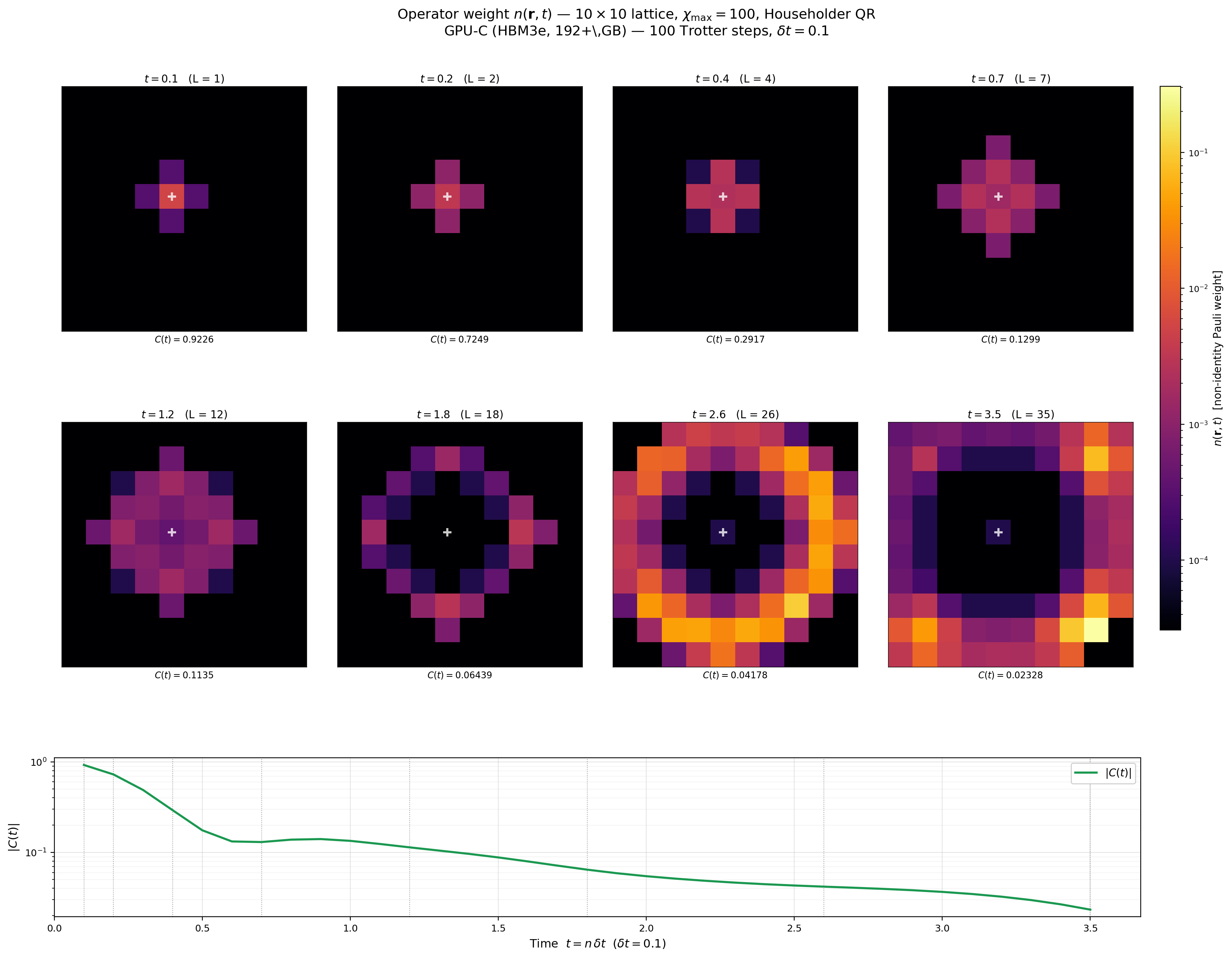}
\caption{Operator weight $n(\mathbf{r},t)$ on the $10\times10$ lattice at
 $\chimax=100$, Householder QR, eight Trotter snapshots (two rows of four,
 $\delta t = 0.1$).
 The operator starts as a single-site excitation at the center $(4,4)$ and
 expands as a Manhattan-distance diamond (Lieb-Robinson cone).
 At $t \approx 2.0$--$2.4$ the expanding front reaches the lattice boundary
 and reflects, forming a bright ring of accumulated weight at sites equidistant
 from the center.
 At later times the ring disperses toward increasingly uniform coverage
 ($S_{\rm frac} \to 1$).
 \emph{Lower panel}: $C(t)$ decays below $10^{-3}$ by $t \approx 2.0$;
 all physical information at late times is carried by $n(\mathbf{r},t)$,
 not by $C(t)$.
 This run requires $\approx 109$\,GB and is enabled by the 192+\,GB configuration.}
\label{fig:lightcone_10x10}
\end{figure}

\begin{figure}[H]
\centering
\includegraphics[width=\linewidth]{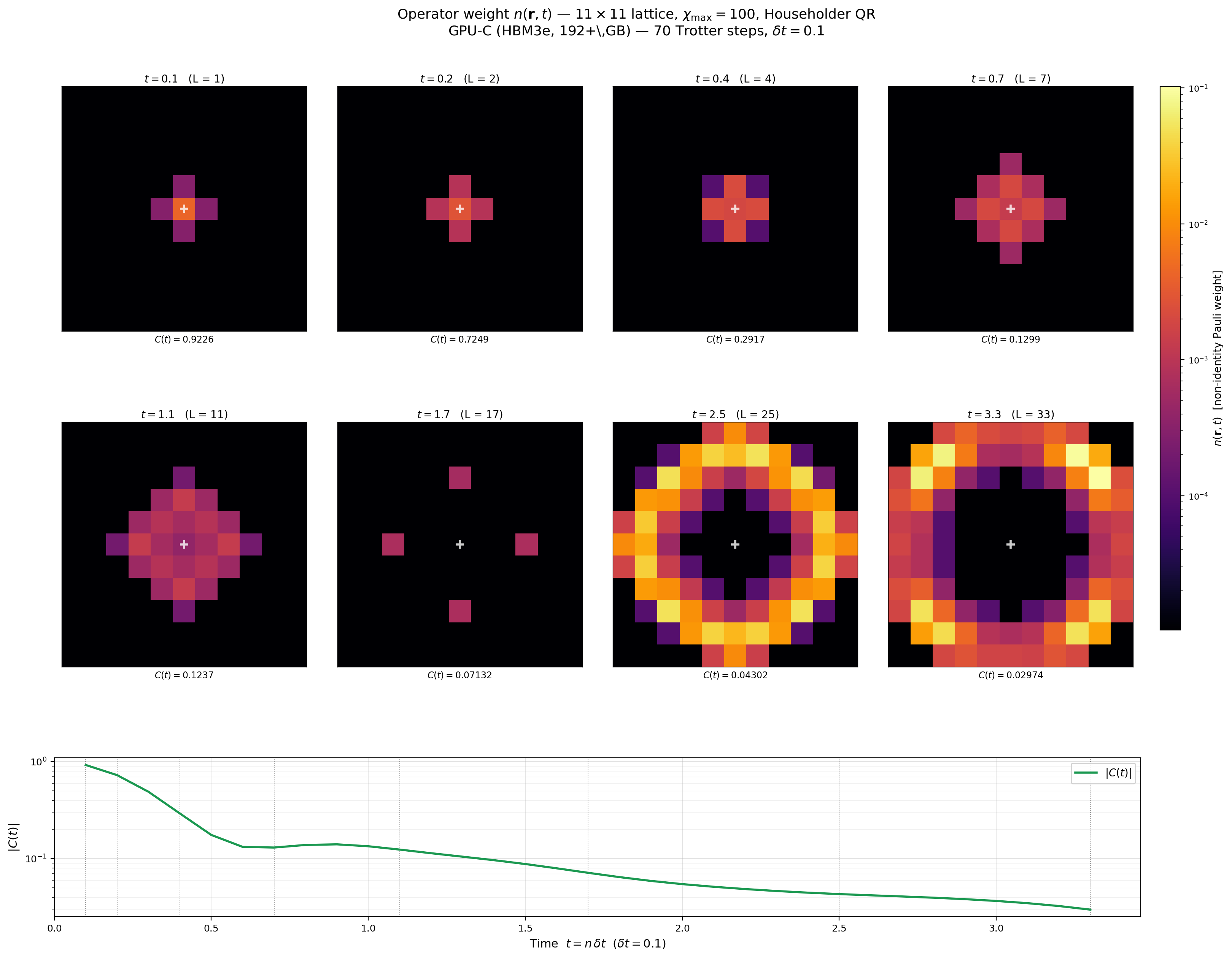}
\caption{Operator weight $n(\mathbf{r},t)$ on the $11\times11$ lattice at
 $\chimax=100$, Householder QR, eight Trotter snapshots (two rows of four).
 The operator starts as a single-site excitation at the center $(5,5)$.
 \textbf{Early phase} ($t \lesssim 0.7$): ballistic expansion as a
 Manhattan-distance diamond (Lieb-Robinson cone).
 \textbf{Boundary reflection} ($t \approx 1.7$--$2.5$): front reaches all
 four edges; reflected waves form a bright ring equidistant from the center.
 \textbf{Scrambling} ($t \gtrsim 2.5$): ring disperses toward uniform coverage.
 The fourfold C4 symmetry is visible at every layer, confirming zero BP trap
 contamination.
 $C(t)$ (lower panel) decays smoothly to ${\approx}0.03$ over 33 layers.
 This run requires $\approx 130$\,GB and is enabled by the 192+\,GB configuration.}
\label{fig:lightcone_11x11_chi100}
\end{figure}

\subsection{Future performance directions}

Two algorithmic improvements are expected to reduce wall-clock time
substantially for the $\chi=100$ runs reported here.

\paragraph{GPU streams and memory-bandwidth overlap.}
The current implementation applies each Trotter layer synchronously:
the SVD singular values are copied device-to-host to determine the
truncation rank before the next contraction can begin, serialising the
GPU pipeline.
With GPU streams, successive Trotter layers can be overlapped ---
the device-to-host transfer for layer $n$ proceeds concurrently with
the tensor contraction of layer $n+1$.
GPU-C's 192+\,GB high-bandwidth memory capacity is the enabling factor: at $\chi=100$
the full state (${\approx}109$\,GB for $10\times10$, ${\approx}130$\,GB
for $11\times11$) fits comfortably in a single device, leaving sufficient
headroom to double-buffer the workspace tensors required for stream overlap.
On GPU-B (32\,GB) this headroom does not exist; streams are therefore
a GPU-C-specific optimisation.
We estimate a $1.5$--$2\times$ reduction in Trotter time per layer once
the SVD synchronisation point is replaced by an on-device reduction.

\paragraph{Float64 gate arithmetic and Adaptive QR robustness.}
The Adaptive QR mode applies Cholesky-QR as its primary factorisation,
falling back to Householder QR when the Gram matrix is ill-conditioned.
Each fallback incurs a penalty --- visible in the BP timing as a
per-layer spike --- because Householder is slower and resets the gauge
environment for subsequent BP sweeps.
Computing the gate (QR, SVD, Cholesky) in float64 while keeping tensor
storage in float32 would improve Cholesky conditioning significantly:
a well-conditioned Gram matrix produces a more uniform gauge across bonds,
which in turn reduces BP sweep count and eliminates most fallback events.
The space cost is modest (only the gate workspace buffers double, not the
full $O(\chi^2)$ state), and the timing cost is expected to be small
because gate computation is arithmetic-bound in the L2-cache regime
at $\chi \leq 100$.
If fallbacks are eliminated, Adaptive QR would run strictly faster than
Householder at $\chi=100$, making it the unambiguous production mode for
GPU-C across all grid sizes.

% content lives in appendix_mi355x.tex

\end{document}